\documentclass[letterpaper,twocolumn,10pt]{article}
\PassOptionsToPackage{hyphens}{url}
\usepackage{hyperref}
\usepackage[hyphenbreaks]{breakurl}
\usepackage{usenix2019_v3}

\usepackage[utf8]{inputenc}
\usepackage[english]{babel}

\usepackage{amsmath}
\usepackage{amsthm} 
\usepackage{amssymb}
\usepackage[inline]{enumitem}
\usepackage{bookmark}
\usepackage{varioref} %
\usepackage{hyperref} %
\usepackage{cleveref} %
\crefname{section}{§}{§§}
\Crefname{section}{§}{§§}
\crefname{subsection}{§}{§§}
\Crefname{subsection}{§}{§§}

\newcommand{\oursystem}{HECO\xspace} %

\usepackage{graphicx}
\usepackage{tabularx}
\usepackage{booktabs}
\usepackage{placeins}
\usepackage{caption}
\usepackage{subcaption}
\usepackage{comment}
\usepackage{enumerate}
\usepackage{xspace}
\def\scalecaption{\normalsize}
\usepackage{multirow}
\usepackage{soul}

\usepackage{breakurl}

\usepackage{enumitem}
\setlist[itemize]{noitemsep, topsep=0pt, leftmargin=10pt}

\usepackage{colortbl}
\definecolor{Gray}{gray}{0.65}
\definecolor{LightGray}{gray}{0.9}

\usepackage{blindtext}

\usepackage[utf8]{inputenc}
\usepackage[english]{babel}
\usepackage{bookmark}
\usepackage{varioref} %
\usepackage{hyperref} %
\usepackage{cleveref} %
\crefname{lstlisting}{listing}{listings}
\Crefname{lstlisting}{Listing}{Listings}

\usepackage{tabularx}
\usepackage{booktabs}
\usepackage{placeins}
\usepackage{caption}
\usepackage{subcaption}
\usepackage{comment}
\usepackage{enumerate}
\usepackage{xspace}
\def\scalecaption{\normalsize}
\usepackage{multirow}

\usepackage{listings,newtxtt}

\usepackage{enumitem}
\setlist[itemize]{noitemsep, topsep=0pt, leftmargin=10pt}

\usepackage{colortbl}
\definecolor{Gray}{gray}{0.65}
\definecolor{LightGray}{gray}{0.9}

\let\oldparagraph=\paragraph
\renewcommand\paragraph[1]{\oldparagraph{#1.}}

\usepackage{bold-extra}
\usepackage[printonlyused,nohyperlinks]{acronym}

\usepackage[binary-units,group-separator={,},range-units=single,range-phrase=--]{siunitx}
\DeclareSIUnit[number-unit-product = {\thinspace}]{\inch}{inch}
\DeclareSIUnit{\bits}{bits}
\DeclareSIUnit{\bit}{bit}
\DeclareSIUnit \px {\ensuremath{\mathit{px}}}

\usepackage{booktabs}

\usepackage{multirow}  	 	

\usepackage{tabularx,ragged2e} 	

\usepackage{mathtools}

\DeclarePairedDelimiterX{\reducedX}[2]{[}{]_{#1}}{#2}

\usepackage{float}
\newfloat{lstfloat}{htbp}{lop}
\floatname{lstfloat}{Listing}
\crefalias{lstfloat}{listing} %
\DeclareCaptionSubType{lstfloat} %
\crefalias{sublstfloat}{listing} %

\usepackage{framed}
\renewenvironment{framed}[1][\hsize]
{\MakeFramed{\hsize#1\advance\hsize-\width \FrameRestore}}%
{\endMakeFramed}

\usepackage{algorithmicx}
\usepackage[noend]{algpseudocode}
\usepackage{algorithm}

\algnewcommand\algorithmiccontinue{\textbf{continue}}
\algnewcommand\Continue{\algorithmiccontinue}
\algnewcommand\algorithmicbreak{\textbf{break}}
\algnewcommand\Break{\algorithmicbreak}

\algnewcommand\algorithmicswitch{\textbf{switch}}
\algnewcommand\algorithmiccase{\textbf{case}}
\algdef{SE}[SWITCH]{Switch}{EndSwitch}[1]{\algorithmicswitch\ #1:}{\algorithmicend\ \algorithmicswitch}%
\algdef{SE}[CASE]{Case}{EndCase}[1]{\algorithmiccase\ #1:}{\algorithmicend\ \algorithmiccase}%
\algtext*{EndSwitch}%
\algtext*{EndCase}%

\algdef{SE}[FORALL]{ForAll}{EndFor}[1]{\algorithmicforall\ #1:}{}%
\algtext*{EndFor}%

\algrenewcommand\algorithmicforall{\textbf{foreach}}

\algnewcommand\FheReplace{\Call{Replace}}
\algnewcommand\FheExtract{\text{fhe.extract}}
\algnewcommand\FheInsert{\text{fhe.insert}}
\algnewcommand\FheRotate{\text{fhe.rotate}}
\algnewcommand\FheSecret{\text{fhe.secret}}
\algnewcommand\FheConst{\text{fhe.ptxt}}
\algnewcommand\FuncReturn{\text{func.return}}
\algnewcommand\Type[1]{\text{type}(#1)}
\algnewcommand\Graph{\mathcal{G}}
\algnewcommand\Vertices{\mathcal{V}}
\algnewcommand\Edges{\mathcal{E}}

\DeclarePairedDelimiterX\Set[1]{\lbrace}{\rbrace}%
{  #1 }

\algrenewcommand\algorithmicfunction{\textbf{Algorithm}}
\algnewcommand\algorithmicassert{\texttt{assert}}
\algnewcommand\Assert[1]{\State \algorithmicassert~#1}%
\algnewcommand{\Downto}{\textbf{ downto }}
\algnewcommand{\By}{\textbf{ by }}

\usepackage{inconsolata}

\definecolor{clr-background}{RGB}{255,255,255}
\definecolor{clr-text}{RGB}{0,0,0}
\definecolor{clr-string}{RGB}{55,125,34}
\definecolor{clr-namespace}{RGB}{0,0,0}
\definecolor{clr-preprocessor}{RGB}{128,128,128}
\definecolor{clr-keyword}{RGB}{0,0,128}
\definecolor{clr-type}{RGB}{204,120,50}
\definecolor{clr-variable}{RGB}{0,0,0}
\definecolor{clr-constant}{RGB}{111,0,138} %
\definecolor{clr-comment}{RGB}{128,128,128}
\definecolor{clr-number}{RGB}{0,0,255}
\definecolor{clr-fhe-types}{RGB}{58,127,194}

\lstdefinestyle{python}{
    language=C++,
    backgroundcolor=\color{clr-background},
    basicstyle=\footnotesize\ttfamily, %
    stringstyle=\color{clr-string},
    identifierstyle=\color{clr-variable}, %
    commentstyle=\color{clr-comment},
    morecomment=[l]{\#},
    keywordstyle=\color{clr-keyword}\bfseries,
    keywordstyle={[2]\color{clr-number}}, %
    keywordstyle={[3]\color{clr-keyword}},
    keywordstyle={[4]\color{clr-fhe-types}}, %
    tabsize=2,
    morekeywords=[3]{range},
    otherkeywords={1, 2, 3, 4, 5, 6, 7, 8, 9, 0},
    morekeywords=[2]{1, 2, 3, 4, 5, 6, 7, 8, 9, 0},
    literate={{r0}{{\color{clr-text}r0}}2 {r1}{{\color{clr-text}r1}}2 {r2}{{\color{clr-text}r2}}2
        {r3}{{\color{clr-text}r3}}2 {r4}{{\color{clr-text}r4}}2 {r5}{{\color{clr-text}r5}}2
        {r6}{{\color{clr-text}r6}}2 {r7}{{\color{clr-text}r7}}2 {r8}{{\color{clr-text}r8}}2
        {sf64}{{\color{clr-text}sf64}}3 {f64}{{\color{clr-text}f64}}3}, %
    deletekeywords={new},
    morekeywords={def, from, import, with},
    frame=single,
    numbers=left,                    %
    numbersep=5pt,                   %
    numberstyle=\tiny\color{gray}, %
    rulecolor=\color{black},         %
    showspaces=false,                %
    showstringspaces=false,          %
    showtabs=false,                  %
    stepnumber=1,                    %
    tabsize=2,	                   %
    title=\lstname,                   %
    showlines=false
    backgroundcolor=\color{white},   %
    breakatwhitespace=false,         %
    breaklines=true,                 %
    captionpos=b,                    %
    deletekeywords={range},            %
    escapeinside={(*@}{@*)},          %
    extendedchars=true,              %
    firstnumber=1                %
}
\lstdefinestyle{mlir}{
    language=C++,
    backgroundcolor=\color{clr-background},
    basicstyle=\footnotesize\ttfamily, %
    stringstyle=\color{clr-string},
    identifierstyle=\color{clr-variable}, %
    commentstyle=\color{clr-comment},
    morecomment=[l]{\#},
    keywordstyle=\color{clr-fhe-types},
    tabsize=2,
    alsoletter={.},
    morekeywords={fhe.extract,fhe.insert,fhe.add,fhe.mul,fhe.rotate},
    otherkeywords={},
    frame=single,
    numbers=left,                    %
    numbersep=5pt,                   %
    numberstyle=\tiny\color{gray}, %
    rulecolor=\color{black},         %
    showspaces=false,                %
    showstringspaces=false,          %
    showtabs=false,                  %
    stepnumber=1,                    %
    tabsize=2,	                   %
    title=\lstname,                   %
    showlines=false
    backgroundcolor=\color{white},   %
    breakatwhitespace=false,         %
    breaklines=true,                 %
    captionpos=b,                    %
    deletekeywords={range},            %
    escapeinside={(*@}{@*)},          %
    extendedchars=true,              %
    firstnumber=1                %
}

\newcommand{\cpp}{C\texttt{++}\xspace}

\DeclareMathOperator{\Enc}{Enc}
\DeclareMathOperator{\Dec}{Dec} 

\usepackage{etoolbox}
\usepackage{color}
\usepackage{algpseudocode}
\begin{document}
\newacro{alchemy}[\textsc{Alchemy}]{A Language and Compiler for Homomorphic Encryption Made easY}
\newacro{ast}[AST]{Abstract Syntax Tree}
\newacro{aes}[AES]{Advanced Encryption Standard}

\newacro{bfv}[BFV]{Brakerski/Fan-Vercauteren}
\newacro{bgv}[BGV]{Brakerski-Gentry-Vaikuntanathan}
\newacro{bfs}[BFS]{Breadth-First Search}

\newacro{cpu}[CPU]{Central Processing Unit}
\newacro{cggi}[CGGI]{Chillotti-Gama-Georgieva-Izabachene}
\newacro{cgt}[CGT]{Class Generation Tool}
\newacro{ckks}[CKKS]{Cheon-Kim-Kim-Song}
\newacro{crt}[CRT]{Chinese Remainder Theorem}
\newacro{cfg}[CFG]{Control Flow Graph}
\newacro{cse}[CSE]{Common Subexpression Elimination}
\newacro{ctr}[CTR]{Click-Through Rate}
\newacro{ccpa}[CCPA]{California Consumer Protection Act}
\newacro{chet}[CHET]{Compiler and Runtime for Homomorphic Evaluation of Tensor Programs}
\newacro{cnn}[CNN]{Convolutional Neural Network}
\newacro{cmux}[CMux]{Conditional Multiplexer}
\newacro{ci}[CI]{Continuous Integration}

\newacro{dsl}[DSL]{Domain-Specific Language}
\newacro{dft}[DFT]{Discrete Fourier Transform}
\newacro{dag}[DAG]{Directed Acyclic Graph}
\newacro{dfs}[DFS]{Depth-First Search}
\newacro{dfg}[DFG]{Data Flow Graph}
\newacro{ml}[ML]{Machine Learning}

\newacro{e3}[E\textsuperscript{3}]{Encrypt-Everything-Everywhere}
\newacro{e2ee}[E2EE]{End-to-End Encryption}
\newacro{eva}[EVA]{Encrypted Vector Arithmetics Language and Compiler}

\newacro{fft}[FFT]{Fast Fourier Transformation}
\newacro{fhe}[FHE]{Fully Homomorphic Encryption}
\newacro{ftt}[FTT]{Fermat-Theoretic Transform}
\newacro{flash}[FLaSH]{Fully, Leveled and Somewhat Homomorphic Encryption Library}

\newacro{gsw}[GSW]{Gentry-Sahai-Waters}
\newacro{gpv}[GPV]{Gentry-Peikert-Vaikuntanathan}
\newacro{gpu}[GPU]{Graphics Processing Unit}
\newacro{gwas}[GWAS]{Genome-Wide Association Studies}
\newacro{gdpr}[GDPR]{General Data Protection Regulation}

\newacro{he}[HE]{Homomorpic Encryption}
\newacro{helib}[HElib]{Homomorphic Encryption Library}
\newacro{heaan}[HEAAN]{Homomorphic Encryption for Arithmetic of Approximate Numbers}

\newacro{ir}[IR]{Intermediate Representation}
\newacro{ide}[IDE]{Integrated Development Environment}

\newacro{json}[JSON]{JavaScript Object Notation}

\newacro{lwe}[LWE]{Learning With Errors}
\newacro{lut}[LUT]{Look-Up Table}

\newacro{mpc}[MPC]{Multi-Party Computation}

\newacro{ntt}[NTT]{Number-Theoretic Transform}

\newacro{pahe}[PAHE]{Packed Additively Homomorphic Encryption}
\newacro{prf}[PRF]{Pseudorandom Function}

\newacro{rtl}[RTL]{Register-Transfer-Level}
\newacro{rns}[RNS]{Residue Number System}
\newacro{rlwe}[RLWE]{Ring-Learning With Errors}

\newacro{s-expression}[s-expression]{Symbolic Expression}
\newacro{seal}[SEAL]{Simple Encrypted Arithmetic Library}
\newacro{simd}[SIMD]{Single Instruction, Multiple Data}
\newacro{stst}[StSt]{Stehle-Steinfeld}
\newacro{she}[SHE]{Somewhat Homomorphic Encryption}
\newacro{ssa}[SSA]{Static Single Assignment}
\newacro{stl}[STL]{Standard Template Library}
\newacro{slp}[SLP]{Superword-Level Parallelism}

\newacro{tfhe}[TFHE]{Fast Fully Homomorphic Encryption Library over the Torus}

\newacro{uml}[UML]{Unified Modeling Language}

\newacro{wfa}[WFA]{Weighted Finite Automata}

\newacro{zkp}[ZKP]{Zero-Knowledge Proofs}

\date{}
\title{\large \bf \oursystem : Fully Homomorphic Encryption Compiler \vspace{-15pt} 
}
{

\author{
{ \vspace{5pt} \rm Alexander Viand, Patrick Jattke, Miro Haller, 
Anwar Hithnawi}  \\ 
{\textit{ETH Zurich}}  %
} 

\graphicspath{{./images/}}

\patchcmd{\maketitle}
{\@maketitle}
{\vspace{-6em}\@maketitle \vspace{-3em}}%
{}
{}

\maketitle

\begin{abstract}
    In recent years, \acf{fhe} has undergone several breakthroughs and advancements leading to a leap in performance.
Today, performance is no longer a major barrier to adoption.
Instead, it is the complexity of developing an \emph{efficient} FHE application that currently limits deploying FHE in practice and at scale.
Several FHE compilers have emerged recently to ease FHE development. 
However, none of these answer how to automatically transform imperative programs to secure and efficient FHE implementations. 
This is a fundamental issue that needs to be addressed before we can realistically expect broader use of FHE. 
Automating these transformations is challenging because
the restrictive set of operations in FHE and their non-intuitive performance characteristics require programs to be drastically transformed to achieve efficiency.
Moreover, existing tools are monolithic and focus on individual optimizations. 
Therefore, they fail to fully address the needs of end-to-end FHE development.
In this paper, we present HECO, a new end-to-end design for FHE compilers that takes high-level imperative programs and emits efficient and secure FHE implementations. 
In our design, we take a broader view of FHE development, extending the scope of optimizations beyond the cryptographic challenges existing tools focus on.

\end{abstract}

\section{Introduction} \label{intro}

Privacy and security are gaining tremendous importance across all organizations, as public perception has shifted and expectations, including regulatory demands, have increased. 
This has led to a surge in demand for secure and confidential computing solutions that protect data's confidentiality in transit, rest, and in-use. 
\acf{fhe} is a key secure computation technology that enables systems to preserve the confidentiality of data at any phase; hence, allowing outsourcing of computations without having to grant access to the data.
In the last decade, theoretical breakthroughs propelled FHE to a practical solution for a wide range of applications~\cite{Kannepalli2021-vh, Kim2020-dk, Carpov2020-gl} in real-world scenarios. 
In addition, end-user facing deployments have started to appear, e.g., in Microsoft Edge's password monitor~\cite{Kannepalli2021-vh}.    
With upcoming hardware accelerators for FHE promising further speedup~\cite{dprive,Samardzic2021-xo}, FHE will soon be competitive for an even wider set of applications.

Though promising in its potential, developing \emph{efficient} FHE applications remains a complex and tedious process that even experts struggle with.
A large part of this complexity arises from the need to map applications to the unique programming paradigms imposed by FHE (cf. \Cref{paradigm}).
Properly optimized code for this paradigm is often several orders of magnitude more efficient than poorly adapted code, making optimization essential for practical FHE applications and posing a major barrier to wider adoption.

Fully Homomorphic Encryption is a nascent field and still actively evolving, with ongoing research on the cryptography, software implementations, and, increasingly, on hardware accelerators.
As a result, tools must be designed to accommodate and adapt to to this fast moving field.
Existing compilers (cf. \Cref{relwork}), however, are mostly rigid, monolithic tools with a narrow focus on individual sub-optimizations.
Whereas experts usually transform applications in ways that accelerate them by orders of magnitude, existing tools have mostly focused on smaller-scale optimizations that result in small constant-factor speedups.
While these represent important contributions, they are insufficient to make the kind of qualitative performance difference that is necessary to achieve practical FHE.
In order to overcome these limitations, we need to fundamentally \emph{rethink the architecture of  FHE compilers} and \emph{develop novel optimizations} that abstract away the complexity FHE and address the limitations of existing tools.
In this paper, we present \oursystem, a new multi-stage optimizing FHE compiler. 
Our architecture provides, for the first time, a true end-to-end toolchain for FHE development. %
In addition, we propose novel transformations and optimizations that map imperative programs to the unique programming model of FHE.%

\paragraph{E2E Architecture}
Expert developers naturally structure the development of FHE applications into different stages.
First, they consider how to efficiently map applications to the unique paradigm of FHE, e.g., how to minimize the need for data movement.
Only afterward do they consider lower-level issues, such as fine-tuning the program to exploit scheme-specific optimizations.
Currently, most developers then target software libraries such as Microsoft SEAL~\cite{sealcrypto} which realize the underlying FHE schemes and implement a range of cryptographic optimizations.
However, with more focus on hardware acceleration, there is an emerging need to optimize for individual hardware targets, which libraries are usually ill-suited to accomplish.
Based on this progression of abstraction levels, 
we identified four phases of converting an application to an efficient FHE implementation: \emph{program transformation}, \emph{circuit optimization}, \emph{cryptographic optimization}, and \emph{target optimization} (cf. \Cref{system-overview}).

Compilers need to be able to accommodate a wide variety of optimizations across these different levels of abstraction.
However, existing compilers usually abstract FHE computations as circuits  consisting of the basic homomorphic operations and scheme-specific operations for managing noise.
This is a natural representation since FHE computations are mathematically modeled as arithmetic circuits.
However, many optimizations at the high-level (program transformation) or at the lowest level (target optimization) cannot be fully expressed in the circuit setting because they consider aspects such as data flow or memory management which have no natural correspondence in this abstraction.
We instead propose a set of \acfp{ir} based on the requirements of each phase that allow us to naturally and efficiently express optimizations at these different levels.
We realize these \acp{ir} using the MLIR compiler framework~\cite{mlir}, which provides a standardized way to define and operate on domain-specific \acp{ir}.
MLIR enables the transfer of optimizations between different projects, including across domains, and provides a powerful software framework.
Specifically, it is well suited to represent and optimize high-level programs in a way that circuit-based tools are not.

\vspace{-10pt}
\paragraph{Automated Mapping to Efficient FHE}
\oursystem supports the automatic transformation of high-level programs to FHE's unique programming paradigm.
Experts spend significant time considering \emph{how} to best express an application in the FHE paradigm, only considering the other aspects once the program is \emph{efficiently} expressible using native FHE operations.
Existing tools, in contrast, typically disregard this arguably most important phase of the FHE development process.

In \oursystem, developers can express their algorithms conveniently in the standard imperative paradigm, e.g., using loops that access/modify individual vector elements.
However, such programs do not align well with the restricted set of operations offered by \ac{fhe} schemes, requiring the compiler to translate and restructure the application.
We focus on transformations targeting the \ac{simd} parallelism of most modern schemes, which allows one to \emph{batch} many (usually, $2^{13}-2^{16}$) different values into a single ciphertext and compute over all simultaneously.
Batching is used by experts to drastically reduce ciphertext expansion and computational overhead, frequently improving runtimes by several orders of magnitude when compared to naive implementations.

While it is arguably the single most important optimization for many applications, unlocking its performance potential currently requires significant expertise and experience in writing FHE applications.
Because these schemes do not offer the data movements operations (e.g., scatter/gather/permute) usually present in the context of vector operations, existing approaches from the traditional compiler literature do not translate well.
Specifically, FHE only natively supports element-wise SIMD operations and cyclical rotations of the elements inside a ciphertext and existing algorithms must be transformed dramatically in order to be expressed solely from these operations.
We devise a series of transformations and optimizations that can translate batching-amenable programs to fully exploit SIMD operations while minimizing the need for data movement (c.f. \Cref{design}).

In our evaluation, we show that HECO can match the performance of expert implementations, providing up to 3500x speedup over naive non-batched implementations (c.f. \Cref{evaluation}).
We open-source \oursystem and we hope that it will help to advance the FHE development ecosystem. 
\oursystem decouples optimizations from front- and back-end logic, allowing it to be easily extended to different languages, FHE libraries,  accelerators, and novel optimizations as they emerge.

\section{Background}
\label{background}

\begin{figure}[t]
    \center
    \includegraphics[width=0.95\columnwidth]{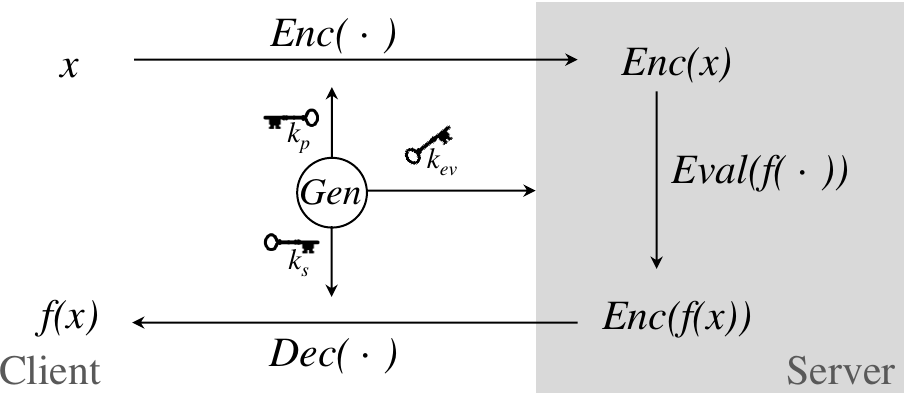}
    \caption{Using FHE, a third party can compute on encrypted values without requiring access to the underlying data.}
    \vspace{-1em}
    \label{fhe_fig}
\end{figure}

In this section, we briefly introduce the notion of \ac{fhe} and key aspects of modern FHE schemes.

\subsection{Fully Homomorphic Encryption}
In a \emph{homomorphic} encryption (HE) scheme, there exists a homomorphism between plaintext and ciphertext operations such that, e.g., $\Dec(\Enc(x+y)) = \Dec(\Enc(x) \oplus \Enc(y))$ in the case of additively homomorphic encryption.
While \emph{additively} and \emph{multiplicative} HE schemes (e.g., Paillier~\cite{Paillier1999a} and textbook RSA~\cite{Rivest1978}, respectively) have been known for many decades, \emph{fully} homomorphic encryption (FHE) schemes that support an arbitrary combination of both operations remained practically infeasible until Gentry's breakthrough in 2009~\cite{Gentry2009}.
This allows a third party to compute on encrypted data,  without requiring access to the underlying data.
Specifically, FHE is traditionally defined for \emph{outsourced computation}, where a client provides encrypted data $x$ and a function $f$ to a server which computes and returns $f(x)$ as shown in \Cref{fhe_fig}.
The client can decrypt the returned result while the server learns nothing about inputs, intermediate values, or results.
Beyond this simple setting, the function (and additional inputs) can also be supplied by the server, opening up additional interesting deployment scenarios.
For example, Private Set Intersection (PSI) as used in Microsoft Edge's password monitor~\cite{Kannepalli2021-vh}, where the client sends encrypted login credentials to the server, which compares them against its own database of leaked usernames and passwords.
In the decade since its first realization, FHE performance has improved dramatically: from around half an hour to compute a single multiplication to only a few milliseconds.
Nevertheless, this is still seven orders of magnitude slower than a standard, signed multiplication executed on a modern CPU.
However, this gap is expected to be reduced significantly with  the emergence of dedicated FHE hardware accelerators~\cite{Samardzic2021-xo, Samardzic2022-mv, Cammarota2022-ru, Geelen2022-tw}.

\vspace{-1em}
\subsection{FHE Schemes}
We briefly describe the \acs{bfv} scheme~\cite{brakerski2012fully,fan2012somewhat} as a representative of the largest family of FHE schemes.
We focus on aspects relevant to FHE application development and refer to the original papers for further details.
Since most modern FHE schemes follow a similar pattern, the descriptions also mostly apply to the \acs{bgv} and \acs{ckks} schemes~\cite{Brakerski2011b, cheon2017homomorphic}.

In BFV, plaintexts are polynomials of degree $n$ (usually $n>2^{12}$) with coefficients modulo $q$ (usually $q > 2^{60}$).
Encryption introduces \emph{noise} into the ciphertext, which is initially small enough to be rounded away during decryption but accumulates during homomorphic operations.
As basic operations, BFV supports additions and multiplications over ciphertexts.
Additions increase the noise negligibly, but multiplications affect it significantly.
Managing noise is crucial to prevent ciphertext corruption, which manifests as a failing decryption.
The noise limits computations to a (parameter-dependent) number of consecutive multiplications (multiplicative \emph{depth}) before decryption fails.  
While \emph{boostrapping} can reduce the noise homomorphically, it introduces significant overheads and must therefore be eliminated or minimized to achieve practical FHE solutions.

Using the \acf{crt}, it is possible to \emph{encode} a vector of $n$ integers into a single polynomial, with addition and multiplication acting slot-wise (SIMD).
Since the polynomial of degree $n$ is usually between $2^{13}$ and $2^{16}$ for security, it can significantly reduce ciphertext expansion and computation cost.
BFV also supports rotation operations over such \emph{batched} ciphertexts, which cyclically rotate the vector's elements.
Finally, BFV includes a variety of noise-management (or \emph{ciphertext maintenance}) operations, which do not change the encrypted message but can reduce noise growth during computations. 

\vspace{-1em}
\paragraph{Security}
Modern FHE schemes rely on post-quantum hardness assumptions, widely believed to be secure for the foreseeable future.
The community has developed estimates of their concrete hardness~\cite{Albrecht2015-zf} and parameter choices for several FHE schemes have been standardized~\cite{HomomorphicEncryptionSecurityStandard}.
Nevertheless, some attention must be paid to security when using FHE.
For example, \ac{fhe} does not provide \emph{integrity} by default, i.e., a server might perform a different calculation than requested or none at all.
There exist techniques to address that, ranging from zero-knowledge-proofs to hardware attestation~\cite{Brenna2021-hg}.
Additionally, \ac{fhe} does not provide by default \emph{circuit privacy}, i.e., a client might be able to learn information about the applied circuit.
Different techniques, varying in practicality and protection level, can be used to address this~\cite{Bourse2016-zu,Ducas2016-va}.
	Finally, issues can appear when using \emph{approximate} homomorphic encryption (e.g., CKKS),
	with attacks that can recover the secret key from the noise embedded in ciphertext decryptions~\cite{Li_undated-qo}.
	Recent work has shown how adding differentially private noise can mitigate these attacks~\cite{Li2022-zf},
	but some concerns remain.

\vspace{-1em}
\subsection{\ac{fhe} Programming Paradigm} \label{paradigm}
\ac{fhe} imposes a variety of restrictions on developing programs: some derive from the definition of \ac{fhe} and its security guarantees, while others result from scheme restrictions and cost models.
For example, FHE's security guarantees make it necessarily data-independent, hence preventing branching based on secret inputs. While some forms of branching can be \emph{emulated}, all branches must be evaluated, resulting in a potentially significant degradation of performance.
In addition, FHE schemes only offer a limited set of data types and operations, with addition and multiplication as basic operations.
Applied over binary plaintext spaces~($\mathbb{Z}_2$), this technically enables arbitrary computation.
However, the best performance is usually achieved with larger plaintext spaces (e.g., $\mathbb{Z}_t$ for $t \gg 2$).
In this setting, computations are equivalent to arithmetic circuits, which can only compute polynomial functions.
Non-polynomial functions can be approximated, but this is typically prohibitively inefficient.
While recent works have explored homomorphic conversions between binary and arithmetic settings~\cite{Boura2018-wj, pegasus} and introduced \emph{programmable bootstrapping} to approximate non-polynomial functions~\cite{Chillotti2020-ia}, these approaches are not yet practical enough for widespread adoption.

As a result, developing FHE applications requires fundamentally rethinking how programs are written.
Generally, developers need to rethink their approach, e.g., using branch-free algorithms well-suited to low-degree polynomial approximations.
In addition, the large size of FHE ciphertexts, which is required for security reasons, is a significant source of both communication and computation overhead. 
However, it also presents an opportunity, as many\footnote{Specifically, schemes from the Ring-LWE family that B/FV, BGV and CKKS belong to.} schemes support \emph{batching}, which allows encrypting many values into the same ciphertext.
This reduces ciphertext expansion and enables element-wise operations in a \acf{simd} fashion.
Data movement in FHE is incredibly restricted, affording no efficient ways to permute the batched data after encryption, with the exception of cyclical rotations.
As a result, efficient FHE algorithms are usually drastically different from their plaintext equivalents.
Adapting to this unique programming paradigm requires a lot of experience and poses a significant barrier to entry for non-experts.

\begin{figure}[t]
	\center
	\includegraphics[width=0.95\columnwidth]{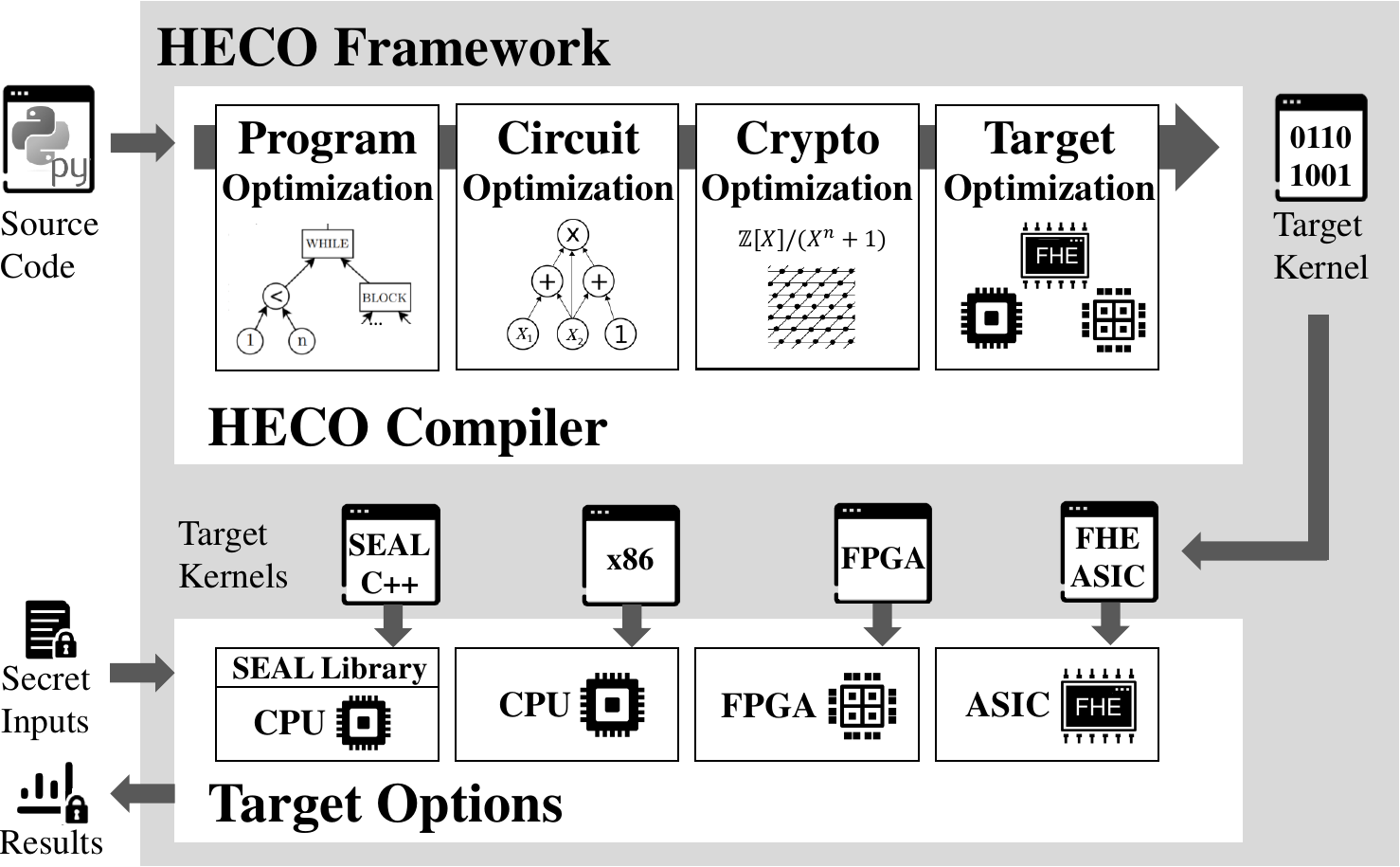}
	\caption{Overview of our end-to-end design, showing the compilation flow from a high-level input program to an efficient FHE kernel running on a target backend.}
	\label{overview_fig}
	\vspace{-1.25em}
\end{figure}

\section{End-to-End FHE Compiler Design}

This section first provides a system overview of \oursystem and then presents its core components,
beginning with the overall framework design, followed by a discussion of our compiler architecture, and finally, give an overview of the transformations and optimizations that constitute the compilation pipeline.

\vspace{-1em}
\subsection{System Overview} \label{system-overview}

\oursystem proposes a multi-staged approach to FHE compilation that encompasses: 
\emph{(i)}~\emph{Program Transformations}, which restructure
high-level programs to be efficiently expressible using native FHE operations,
\emph{(ii)}~\emph{Circuit Optimizations}, which primarily focuses on changes that reduce noise growth in the FHE computation, \emph{(iii)}~\emph{Cryptographic Optimizations}, which instantiate the underlying scheme as efficiently as possible for the given program,
and \emph{(iv)}~\emph{Target Optimizations}, which map the computation to the capabilities of the target.
We propose a set of \acfp{ir} designed to provide a suitable abstraction of each stage, allowing us to naturally and efficiently express optimizations at these different levels.
In contrast, existing compilers usually abstract FHE computations as circuits\footnote{This is natural since FHE computations are usually modeled mathematically as arithmetic circuits.} which does not allow them to fully express many optimizations at the high-level (program transformation) or at the lowest level (target optimization) because these need to consider aspects such as data-flow or memory management which have no natural correspondence in a circuit representation.
In \oursystem, high-level programs are lowered through  a series of transformations, using multiple increasingly lower-level \acp{ir} to produce the target kernel.
These kernels can then be targeted and run against various back-end options.
We provide a user-facing Python framework that abstracts away the complexities of this process, supports a Python-embedded Domain Specific Language for FHE, and provides a unified experience for development, compilation and execution. 
We provide an overview of our end-to-end design in \Cref{overview_fig}.
In the remainder of this section, we describe \oursystem's components, abstractions, and compilation stages.

\subsection{HECO Framework}

\begin{lstfloat}[t]
	\centering
\begin{lstlisting}
def server(x_enc, y_enc, public_context):
  p = FrontendProgram()
    with CodeContext(p):
      def euclidean_sq(x: Tensor[8, Secret[int]],
                       y: Tensor[8, Secret[int]])
                      -> Secret[int]:
        sum: Secret[int] = 0
        for i in range(8):
          d = x[i] - y[i]
          sum = sum + (d * d)
        return sum
    
  # compile FHE code
  f = p.compile(context=public_context)	
  # run FHE code using SEAL
  r_enc = f(x_enc, y_enc)
  return r_enc\end{lstlisting}
	\vspace{-2.5em}
	\caption{Example server-side code using HECO.}
	\label{listing:frontend:server}
\end{lstfloat}

\begin{lstfloat}[t]
    \centering
    \begin{lstlisting}
def client(x : Tensor[int], y : Tensor[int]):
   # Select SEAL backend, scheme and params
   context = SEAL.BFV.new(poly_mod_degree=2048)

   # encrypt input	
   x_enc = context.encrypt(x)
   y_enc = context.encrypt(y) 

   # send enc input to server
   r_enc = server(x_enc, y_enc, context.pub())
   result = context.decrypt(r_enc, context)	
\end{lstlisting}
    \vspace{-1.25em}
    \caption{Corresponding client-side code, outsourcing the computation of  the (squared) euclidean distance.}
    \vspace{-1.25em}
    \label{listing:frontend:client}
\end{lstfloat}

\oursystem's framework ties together the front end, compiler, and the various back ends into a unified development experience. It allows developers to edit, compile and deploy their applications from a familiar Python environment.
	In order to provide an intuition of the developer experience in our system, we provide an example of using \oursystem to compile and run an FHE program in \Cref{listing:frontend:server} (Server) and \Cref{listing:frontend:client} (Client). 
	By wrapping FHE functions in \texttt{with} blocks, we can operate on them as first-class entities, making compilation explicit.
	Our framework provides the necessary infrastructure to run programs directly from the front end,
	allowing developers to integrate FHE functionality into larger applications easily.

\paragraph{Python-Embedded DSL}
	\oursystem uses Python to host its \acf{dsl}, inheriting Python's syntax and general semantics. 
	We want to allow developers to write programs in as natural a fashion as possible, and merely require type annotations to denote inputs that are \texttt{Secret}.
	In order to facilitate this,
	\oursystem supports (statically sized) loops, access to vector elements, and many other high-level features that do not have a direct correspondence in FHE.
	Since our compilation approach requires a high-level representation of the input program, including these non-native operations and the control-flow structure, we cannot follow the approach used by most existing tools.
	These tend to execute the program using placeholder objects that record operations performed on them, which is equivalent if considering FHE programs as circuits but removes most of the high-level information about the program structure.
	Instead, we use Python's extensive introspection features to parse the input program and translate the resulting \acf{ast} directly to our high-level \ac{ir}. %

\FloatBarrier

\subsection{Compiler Infrastructure}
	The core of \oursystem is an optimizing compiler that translates and optimizes programs by lowering them through a series of progressively lower-level \acfp{ir}.
	This section describes how we build upon the MLIR framework to realize \oursystem's compiler design.

\paragraph{Multi-Level Intermediate Representations}
	
	HECO's middle end exposes multiple levels of abstractions 
	to facilitate our multi-stage compilation \& optimization approach.
	This is realized through a series of \acfp{ir},	as seen in \Cref{dialects}.
	We leverage the MLIR framework~\cite{mlir}, which was designed specifically to facilitate \emph{progressive lowering}, introducing additional IRs to reduce the complexity of each lowering step.
MLIR defines a common syntax for IR operations, %
for example, an addition might be represented as \texttt{\%2~=~artih.addi(\%0,~\%1)~:~(i16,~i16)~->~i16}.
MLIR is strongly typed,
 however, for conciseness, we will omit the details of type conversions when discussing transformations.
Intermediate Representations in MLIR are composed of sets of operations known as \emph{dialects}.
We define a custom dialect for our high-level abstraction of FHE (\texttt{heco::fhe}) and combine this with built-in dialects for vector operations (\texttt{mlir::tensor}), plaintext arithmetic (\texttt{mlir::arithmetic}) and basic program structure (\texttt{mlir::affine}, \texttt{mlir::func}) to realize our High-Level Intermediate Representation (HIR).
In addition, we define dialects for each of the supported FHE schemes, mirroring their natively supported operations (\texttt{heco::bfv}, \texttt{heco::bgv}, \texttt{heco::ckks}).
MLIR also includes a variety of standard simplification passes, which can be extended to custom dialects by defining appropriate interfaces.

\paragraph{Supporting Different Back-Ends}

FHE is actively evolving, and as such, tools need to be able to adapt to new and improved implementations, both in software and hardware.
This requires a high level of modularity and flexibility from the compiler.
In \oursystem, we achieve this by using target-specific dialects, which can be customized and extended as new back ends are introduced.
While traditional library-based implementations targeting CPUs and GPUs share a common API (conceptually, if not technically), upcoming FPGA and ASIC accelerators for FHE~\cite{dprive, F1, F2, Basa}) feature a much lower-level interface.
These systems are designed to efficiently realize the required mathematical operations in the modular rings of polynomials that underly most FHE schemes, and as a result, their Instruction Set Architectures (ISA) operate on this level.
In order to support this, \oursystem is designed to be easily extended to match this abstraction level,
featuring MLIR dialects for both bignum polynomial ring operations (\texttt{heco::poly}) and for the commonly used \acf{rns} approach using the \acf{crt} to split these large datatypes into hardware-sized elements (\texttt{heco::rns}).
In addition to targeting hardware accelerators, the ability to lower to this level also allows targeting \texttt{x86} directly via LLVM IR and the LLVM toolchain.

\begin{figure}[t]
	\center
	\includegraphics[width=0.99\columnwidth]{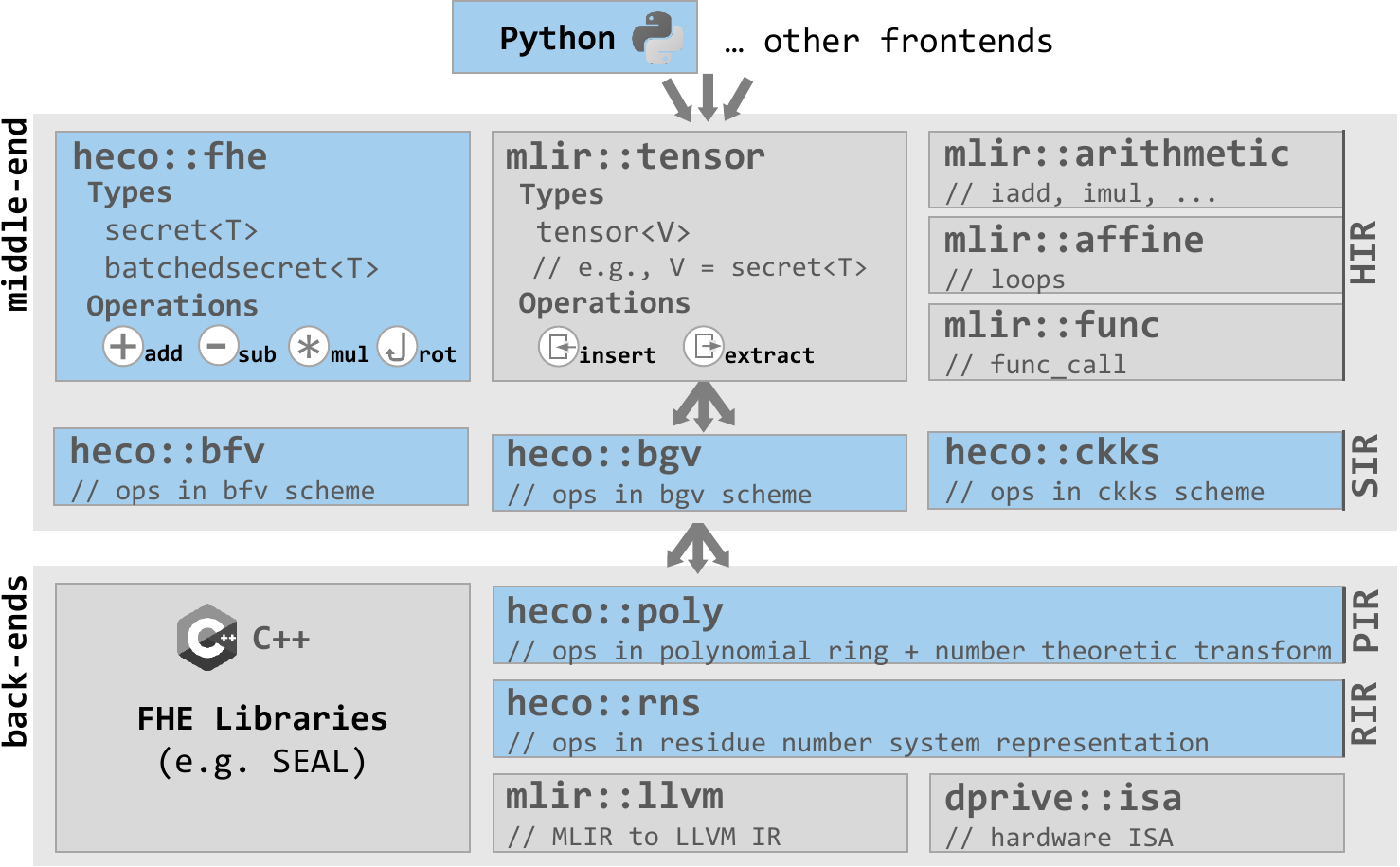}
	\caption{
		Overview of \oursystem's dialects, which define the operations used in the Intermediate Representations (IRs).
	}
	\vspace{-1em}
	\label{dialects}
\end{figure}

\subsection{Transformation \& Optimization}
The transformations and optimizations in \oursystem are grouped according to the four stages of compilation we identified, and we present them accordingly in the following.

\paragraph{Program Transformations}
The first phase of compilation focuses on high-level transformations and optimizations.
This includes a wide variety of general (e.g., constant folding, common sub-expression elimination) and FHE-specific optimizations that allow developers to write code more naturally by removing the need for menial hand-optimization.
Most importantly, however, it focuses on optimizations that map the input program to FHE's unique programming paradigm, such as the automated batching optimizations, which we present in more detail in \Cref{design}.
Previous work has shown that performance differences between the runtimes of well-mapped and naively-mapped implementations can easily reach several orders of magnitude~\cite{viand2021sok}.
As a result, a significant part of our focus in \oursystem is on this level of abstraction, which existing FHE tools generally do not support.

\vspace{-1em}
\paragraph{Circuit Optimizations}
After mapping to the FHE paradigm, the program is conceptually equivalent to an arithmetic circuit of native FHE operations.
This is the level of abstraction considered by the vast majority of existing tools.
Optimizations at this stage are mostly concerned with managing the noise growth in the computation.
For example, a variety of optimizations that try to re-arrange the arithmetic operations to reduce the number of sequential multiplications have been proposed~\cite{Archer2019-iy, Carpov2015-ok, Carpov2018-xi, Aubry2020-jy}.
However, even state-of-the-art optimizations in this style are likely to accelerate a program by only around 2x in practice~\cite{viand2021sok}.
We omit a detailed description of these techniques here as they are not the focus of this paper.
More importantly, this level of abstraction is also where we must consider ciphertext maintenance operations.
These do not modify the encrypted messages, but significantly affect future noise growth, making them essential for practical FHE.
\oursystem uses a traditional approach of inserting relinearization operations~\cite{ fan2012somewhat} between all consecutive multiplications.
This is always correct but not necessarily strictly optimal, and more sophisticated strategies~\cite{Chowdhary2021-go, Dathathri2019-vu, hecate} could offer further improvements.
The modularity of our design makes it a straightforward future work to include these techniques, but since these have been explored in the past, we do not focus on them in this paper.

\begin{lstfloat*}[t]
	\centering
	\begin{sublstfloat}[t]{0.5\textwidth}
		\begin{lstlisting}
def foo(img: Tensor[N, Secret[f64]]):
  img_out = img.copy()
  w = [[1, 1, 1], [1, -8, 1], [1, 1, 1]]
  for x in range(n): # loop over pixels
    for y in range(n):
      t = 0
      for j in range(-1, 2): # apply kernel
        for i in range(-1, 2):
          t += w[i+1][j+1] * img[((x+i)*n+(y+j))%
          img_out[(x*n+y)%
  return img_out\end{lstlisting}
		\vspace{-2.5em}
		\caption{Textbook implementation of a simple sharpening filter}
		\label{batching-example-a}
	\end{sublstfloat}%
	\hfill
	\begin{sublstfloat}[t]{0.45\textwidth}
		\begin{lstlisting}
def foo_batched(img: BatchedSecret[f64]):
  r0 = img * -8
  r1 = img << -n-1
  r2 = img << -n
  r3 = img << -n+1
  r4 = img << -1
  r5 = img << 1
  r6 = img << n-1
  r7 = img << n
  r8 = img << n+1
  return 2*img-(r0+r1+r2+r3+r4+r5+r6+r7+r8)\end{lstlisting}
		\vspace{-2.5em}
		\caption{Optimized batched solution of the same program}
		\label{batching-example-b}
	\end{sublstfloat}
	\caption{
		Both of these functions apply a simple sharpening filter to an encrypted image of size $n \times n = N$,
		by convolving a $3 \times 3$ kernel ($-8$ in the center, $1$ everywhere else) with the image.
		The version on the left encrypts each pixel individually, and follows the textbook version of the algorithm, operating over a vector of $N$ ciphertexts.
		The version on the right batches all pixels into a single ciphertext and uses rotations (\texttt{<\nobreak<}) and \ac{simd} operations to compute the kernel over the entire image at the same time.
		Designing batched implementations requires out-of-the-box thinking in addition to significant expertise and experience.
	}
	\label{batching-example}
	
\end{lstfloat*}

\vspace{-1em}
\paragraph{Cryptographic Optimizations}
In the third phase, we consider \emph{cryptographic optimizations} focused on instantiating the underlying FHE scheme as efficiently as possible. 
When targeting existing FHE libraries, the primary challenge is parameter selection: identifying the smallest (i.e., most efficient) parameters that still provide sufficient noise capacity to perform the computation correctly.
Different techniques have been proposed to estimate the expected noise growth of an FHE program~\cite{noise,Iliashenko2019-ki, precision-loss}.
These include theoretical noise analysis, where recent work has achieved tighter bounds for some schemes~\cite{precision-loss},  but which generally tend to significantly overestimate noise growth~\cite{noise-heuristic-analysis}, leading to unnecessarily large parameter choices.
As a result, experts primarily still rely on a trial-and-error process to experimentally determine the point at which noise invalidates the results.
\oursystem includes basic automatic parameter selection based on a simple multi-depth heuristic but also allows experts to easily override these suggestions.
When %
 targeting hardware directly, rather than through libraries, further optimization opportunities open up.
For example, many ciphertext maintenance operations can be instantiated in different ways, offering trade-offs between runtime, memory consumption, and noise behavior. 
While libraries tend to implement a general-purpose compromise, compilers can adaptively choose the most appropriate approach for a given computation.
However, this requires re-expressing the complex underlying logic of FHE schemes inside the compiler.
\oursystem inherits a powerful system of abstractions and optimizations for computationally intense mathematics from the MLIR framework, allowing our system to be easily extended with such optimizations in the future.
\vspace{-1em}
\paragraph{Target Optimization}
Finally, in the fourth phase, we consider \emph{target-specific optimizations}.
In addition to general code generation optimizations, there is a significant opportunity for FHE-specific optimizations at this level.
For example, when available, FHE benefits greatly from instruction set extensions such as \texttt{AVX512}.
This concept has already been explored in the context of libraries~\cite{hexl}, and implementing similar techniques in \oursystem should be straightforward given our modular design.
When targeting hardware, FHE accelerators impose non-trivial constraints on memory and register usage,
due to complex memory hierarchies.
Initial work in this space has already shown that code generation and scheduling can have a significant impact on accelerator performance~\cite{Basa, F2}.
\oursystem supports optimizations at this level through our low-level dialects for the underlying math, and can easily be extended with target-specific dialects for the Instruction Set Architectures (ISAs) of upcoming accelerators, e.g., those developed by the DARPA DPRIVE program~\cite{dprive}.

\vspace{-1em}

\section{Automatic SIMD Batching} \label{design}

 	In this section, we introduce our automated SIMD batching optimization, which is part of our program transformation stage and maps traditional imperative programs to the restrictive SIMD-like setting of state-of-the-art FHE schemes.

	\vspace{-1em}
\subsection{SIMD Batching}	

Effective use of batching is arguably the single most important optimization for many applications
and is omnipresent in most state-of-the-art FHE results.
Due to the large capacity of FHE ciphertexts (usually $2^{13}-2^{16}$ slots), applying batching has the potential to drastically reduce ciphertext expansion overhead and computation time.
While batching can be used to trivially increase throughput, most FHE applications are constrained by latency.
However, employing batching effectively to improve performance on a single input is non-trivial due to the restrictions imposed by FHE's unusual programming paradigm.
Therefore, unlocking the performance potential of batching currently requires significant expertise and experience in writing FHE applications. In the following, 
	we present a simple example that showcases the drastic transformations that can be required to achieve efficient batching, 
	followed by a brief introduction of a folklore technique that demonstrates common patterns of FHE batching optimizations.

\vspace{-1em}
\paragraph{Example Application -- Image Processing}

	We consider a simple image processing application (see \Cref{batching-example-a}),  which nevertheless features a complex loop nest structure and non-trivial index patterns.
	Specifically, we consider a Laplacian Sharpening filter, i.e., a convolution of a ($3x3$) kernel over an image, implemented with wrap-around padding.
	The function is compatible with efficient arithmetic-circuit based FHE,  as it does not use data-dependent branching and only requires homomorphic addition and multiplications operations.
	However, its current form makes use of nested loops accessing a complex set of indices, which is not very amenable to efficient batching as there appears to be little opportunity for operations over entire ciphertexts.

	Nevertheless, there exists a significantly more efficient \emph{batched} design, as seen in \Cref{batching-example-b}.
	In the optimized version, the input image is batched into a single ciphertext, and all homomorphic operations make full use of their SIMD nature.
	Instead of iterating the kernel over the image, nine copies of the image are made and each is rotated so that all elements interacting at a specific kernel position align at the same index.
	This is possible, because the relative offset between different pixels in the kernel remains static, even though the indices themselves are different for each iteration.
	The transformation enables the the runtime of the program to depend on the (small) kernel size, rather than the image size.
	As a result, the batched version is more than an order of magnitude more efficient than a naive implementation.
	These types of drastic transformations are common in state-of-the-art FHE applications and significant experience is required to develop an intuition for this unusual programming paradigm.

\vspace{-1em}
\paragraph{Rotate-and-Sum}

In the example above, only interactions between values in different ciphertexts were required.
	However, it is also possible to efficiently realize certain operations on the elements of a single ciphertext; we now describe a common folklore technique used to achieve this:
	The \emph{rotate-and-sum} algorithm allows us to efficiently sum up the elements of a ciphertext, using $\O(\log n)$ rotations (where $n$ is the number of ciphertext slots).
	The algorithm proceeds by creating a copy of the current vector, rotating it and then adding both before repeating the same procedure with a lower offset (c.f. \Cref{algo:rotate_and_sum}).
	This is visualized in \Cref{sum-and-rotate} for a vector size of four.
	While this technique is applicable, the performance benefits are usually overshadowed by more radical transformations, such as the example shown above.
	However, using rotate-and-sum and similar rotation-based approaches can be worthwhile if it enables other parts of the program to remain slot-aligned.

\begin{algorithm}[t]
    \begin{algorithmic}[1]
    { \fontfamily{lmr} \selectfont

        \Function{SumVectorPowerTwo}{$x, n$}
            \For{$i \gets \frac{n}{2} \Downto 1 \By i \gets \frac{i}{2}$}
                \State $x \gets x + \Call{Rotate}{x, i}$
            \EndFor
            \Return $x$
        \EndFunction
    }
    \end{algorithmic}
    \caption{Rotate-and-Sum}
    \label{algo:rotate_and_sum}
\end{algorithm}

\begin{figure}[t]
	\center
	\includegraphics[width=1\columnwidth]{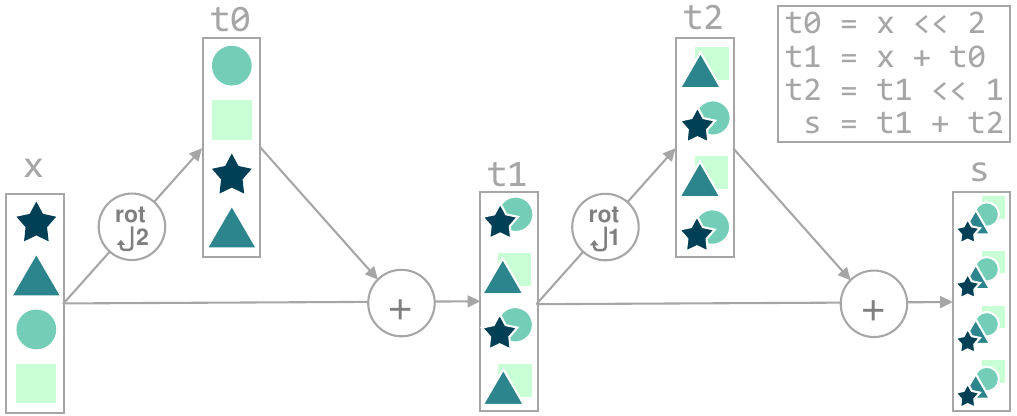}
	\caption{
		Illustration of how repeated copying and rotating can be used to compute the sum of all elements in a ciphertext in a logarithmic, rather than linear, number of steps.
	}
	\label{folding-pattern}
	\label{sum-and-rotate}
\end{figure}

\vspace{-1em}
\subsection{Automatic Batching Approach}

	Experts generally rely on their experience with the FHE programming paradigm 
	to transform and optimize programs for batching, posing a high barrier to entry for non-expert developers.
	Instead, formal methods to automatically translate traditional imperative programs
	into efficient batched FHE solutions are required.
	We assume the input program computes the elements from vectors of secret values in a non-SIMD fashion (e.g., \Cref{batching-example-a}).
	Of course, this can be naively realized by encrypting each vector element into one ciphertext
	, but this usually does not achieve acceptable performance due to the high overhead of FHE.
	The goal of automated batching is to amortize the cost of each FHE operation by utilizing as many ciphertext slots as possible for meaningful computation.
	In the following, we discuss two potential alternative approaches and their drawbacks
	before introducing our approach.

\vspace{-1em}
\paragraph{Strawman Approach}

	Batching each vector in the input program into a ciphertext will trivially achieve a `batched' solution.
	However, this raises the question of how to execute the computations over individual elements present in the program.
	Element-wise access (\texttt{extract} and \texttt{insert}) are not native FHE operations and must instead be emulated, requiring several rotations and ciphertext-plaintext multiplications.
	For example, \Cref{naive-batching-example} shows how \texttt{x[i] = x[i] + y[j]} can be emulated in the batched setting.
	However, this replaces each FHE operation from the naive, vector-of-ciphertexts approach,
	with multiple expensive FHE operations.
	As a result, unless ciphertext expansion is significantly more important than runtime,
	this	approach is virtually always ill-advised.
	Therefore, most existing FHE tools use the vector-of-ciphertexts approach rather than attempting to perform batching.

\begin{lstfloat}[t]
    \begin{framed}[1.03\columnwidth]
    \centering
    \vspace{-5pt}
    \begin{sublstfloat}[t]{0.53\textwidth}
        \begin{lstlisting}[style=mlir,frame=none,otherkeywords={},showlines=true]
%1=tensor.extract %x[i]
%2=tensor.extract %y[j]
%3=fhe.add(%1, %2)
%4=tensor.insert(%3,%z[i])




\end{lstlisting}
        \vspace{-20pt} 
        \caption{Input Program}
        \label{naive-batching-example-a}
    \end{sublstfloat}%
    \hfill
    \begin{sublstfloat}[t]{0.46\textwidth}
        \begin{lstlisting}[style=mlir,numbers=none,frame=none]
%1=fhe.mul(%x, %m_i)
%2=fhe.rotate(%1, -i)
%3=fhe.mul(%y, %m_j)
%4=fhe.rotate(%3, -j)
%5=fhe.add(%2, %4)
%6=fhe.rotate(%5, i)
%7=fhe.mul(%z, %mn_i)
%8=fhe.add(%6, %7)\end{lstlisting}
        \vspace{-20pt} 
        \caption{Naive Batching}
        \label{naive-batching-example-b}
    \end{sublstfloat}
    \end{framed}

    \caption{Strawman batching approach for \texttt{z[i] = x[i] + y[j]}, showing the necessary rotations and multiplications with masking vectors: \texttt{\%m\_i},  \texttt{\%m\_j} are zero everywhere except at $i$ or $j$, respectively; \texttt{\%mn\_i} is one everywhere except at $i$.
    }
    \label{naive-batching-example}
\end{lstfloat}

\vspace{-1em}
\paragraph{Alternative Approaches}
There have been initial attempts at performing automated batching for FHE using Synthesis-based approaches~\cite{Cowan2021-vx}.
While these can, in theory, achieve the drastic transformations required to exploit batching,
they are not suitable for practical use in real-world code development, as they do not scale beyond toy-sized program snippets,	and even those can take minutes to optimize.
Alternatively, one might consider applying traditional \acf{slp} vectorization algorithms~\cite{Larsen2000-xu,Mendis2018-yl, VeGen}, as these try to group operations into SIMD instructions.
However, these generally rely on the ability to efficiently scatter/gather elements into and out of vectors, which is only possible at a high cost in FHE.
While some recent work can reason about the cost of data movement, it does not consider how data movement introduced at the beginning of the program might affect later parts of the program~\cite{VeGen}.

\vspace{-1em}
\paragraph{\oursystem's Approach}

\oursystem's batching transformation starts with the core idea of the strawman approach, 
i.e., batching vectors of secrets into ciphertexts but eliminates the overhead of emulating \texttt{insert}/\texttt{extract} operations for batching amenable programs.
Rather than directly emulating these operations, we instead translate  the homomorphic operations in which they appear as operands.  
While this still requires inserting rotation operations, it allows 
 operations with compatible index access patterns to be mapped to the same emulated code if using an appropriate algorithm.
As a result, a simple built-in simplification pass can eliminate the duplicates.
In the case of well-structured programs, this can completely eliminate emulation related code.
For example, for the program from the previous section (\Cref{batching-example-a}), \oursystem produces exactly the optimized code seen in \Cref{batching-example-b}, which does not contain any emulation-related code.

\vspace{-1em}
\paragraph{HECO Batching Pipeline}

\oursystem's batching transformation is composed of a series of smaller passes, each interleaved with built-in simplification passes.
	Before the main batching passes, we first perform a series of preprocessing steps that unroll statically-sized loops, merge sequential associative binary operations into $n$-ary group operations, and perform a type conversion from vectors of secrets to \texttt{BatchedSecret}s, which are \oursystem's high-level abstraction of ciphertexts.
	Following this, the main pass walks through the program and transforms each operation over secret vector elements into operations over entire vectors. 
	Similar to the strawman approach, this involves introducing rotations, but our approach does not require multiplications with masks.
	Additionally, the way we perform these translations allows us to `chain' them so that consecutive operations on the same vector elements do not result in separate emulation code.
	After the main pass and the associated simplification pass,
	we apply the rotate-and-sum technique where applicable, which is enabled by both the merging of operations during the preprocessing phase and the exposure of same-ciphertext operations by the main pass.
	In the following, we first outline some key preprocessing steps before explaining the two main optimization steps.

\subsection{Preprocessing}
In addition to standard simplifications and canonicalization of the IR (i.e., bringing operations into a standardized `canonical' form to reduce the complexity of the IR), 
we also apply two more specialized transformations.

\paragraph{Merging Arithmetic Operations}
\label{merging}
During the preprocessing stage, we combine chained applications of (associative and commutative) binary operations into larger arithmetic operations with multiple operands (e.g., merging \texttt{x=a+b; y=x+c} to \texttt{y=a+b+c}).
This  removes chains of dependencies and replaces them with a single operation,
making it easier to identify \emph{rotate-and-sum} optimization opportunities.
In addition, it can also allow more efficient direct lowerings, such as when performing the product over $n$ elements:
which can be lowered efficiently to a multiplication tree with depth $\log n$.

\paragraph{Type Conversion}
While tensors can be arbitrarily (re)shaped, all FHE ciphertexts used in a homomorphic computation must have compatible parameters, which implies a fixed number of slots.
	Therefore, we perform type conversion, converting all vector operations over secret values to operations over the \texttt{BatchedSecret} type, an abstract representation of ciphertexts.
	We convert multi-dimensional tensors to vectors using column-major encoding and scale up any secret vector operands to the size of the largest (secret) vector present, padding the plaintext as necessary.
	This does not impact the result of the computation, as the existing code will never access these additional elements.

\subsection{Automatic SIMD-fication}

\begin{algorithm}[t]
	\begin{algorithmic}[1]
		{ \fontfamily{lmr} \selectfont
			
			\Function{BatchPass}{$\Graph$}
			\State $\Vertices, \Edges \gets \Graph$
			\ForAll{$op \in \Vertices \land \Type{op} = \FheSecret$}
			\State $ts \gets \Call{SelectTargetSlot}{op, \Vertices, \Edges}$
			\State \Call{OperandConversion}{$op, ts, \Vertices, \Edges$}
			\ForAll{$v \in \Vertices \land (op, v) \in \Edges$}
			\State $u \gets \FheExtract[v, ts]$
			\State \FheReplace{$v, u, \Vertices, \Edges$}
			\EndFor
			\EndFor
			\EndFunction
			\Procedure{SelectTargetSlot}{$op, \Vertices, \Edges$}
			\ForAll{$v \in \Vertices \land (op, v) \in \Edges$}
			\Switch{$v$}
			\Case{\FheInsert$[\_, i]$}
			\Return $i$
			\EndCase
			\Case{\FuncReturn}
			\Return $0$
			\EndCase
			\EndSwitch
			\EndFor
			\ForAll{$v \in \Vertices \land (v, op) \in \Edges$}
			\Switch{$o$}
			\Case{\FheExtract$[\_, i]$}
			\State \Return $i$
			\EndCase
			\EndSwitch
			\EndFor
			\State \Return $\bot$
			\EndProcedure
			\Procedure{OperandConversion}{$op, ts, \Vertices, \Edges$}
			\ForAll{$v \in \Vertices \land (v, op) \in \Edges \land \Type{v} = \FheSecret $} %
			\Switch{$v$}
			\Case{$\FheExtract(x, i)$}
			\State $u \gets \FheRotate(x, i - ts)$
			\State \FheReplace{$v, u, \Vertices, \Edges$}
			\EndCase
			\Case{$\FheConst[p]$}
			\State $p' \gets \Call{Repeat}{p}$
			\State $u \gets \FheConst(p')$
			\State \FheReplace{$v, u, \Vertices, \Edges$}
			\EndCase
			\EndSwitch
			\EndFor
			\EndProcedure
			\Procedure{Replace}{$v, u, \Vertices, \Edges$} %
			\State $\Vertices \gets \left(\Vertices \setminus \left\{v\right\}\right) \bigcup \left\{u\right\}$
			\ForAll{$w \in \Vertices \land (v, w) \in \Edges$}
			\State $\Edges \gets \left(\Edges \setminus \left\{(v, w)\right\}\right) \bigcup \left\{(u, w)\right\}$
			\EndFor
			\ForAll{$w \in \Vertices \land (w, v) \in \Edges$}
			\State $\Edges \gets \left(\Edges \setminus \left\{(w, v)\right\}\right) \bigcup \left\{(w, u)\right\}$
			\EndFor
			\EndProcedure
		}
	\end{algorithmic}
	\caption{Batching Pass}
	\label{algo:batching_pass}
\end{algorithm}

This pass replaces scalar operations over vector elements (e.g., \texttt{x[i] + y[j]}) with SIMD operations, applying the same operation to each element of the ciphertext.
At its core, the pass is a linear walk over all (arithmetic) homomorphic operations in the program, as seen in \Cref{algo:batching_pass}.
For each operation, we \textit{(i)} identify in which ciphertext slot the result should be computed \textit{(ii)} transform the operands so that they are suitable for such a SIMD-operation, and \textit{(iii)} insert \texttt{extract} operations in situations where a scalar is expected (including uses in later operations that have yet to be transformed).
Note that, by itself, this transformation does not actually remove any code. 
However, it will expose common patterns (e.g., such as those occurring in loops) and cause operations over \emph{compatible} indices to be translated to the same SIMD operations.
This allows the following clean-up pass to remove these now-redundant operations, 
which frequently includes duplicate arithmetic operations and many or all of the operations inserted to ensure consistency.

\paragraph{Target Slot Selection}

	When translating an operation with operands corresponding to different vector positions (e.g., \texttt{x[i]+y[j]}), we must bring the elements of interest into alignment by issuing a rotation for at least some of the operands.
	However, there are usually multiple valid solutions (e.g., \texttt{rot(x, j-i)} vs. \texttt{rot(y, i-j)}), especially for operations with multiple operands, which occur frequently as a result of our preprocessing stage
An obvious approach would be to rotate each operand (e.g., \texttt{x[i]}) so that the element of interest is moved to slot 0.
Then, performing the SIMD operation over the rotated operands produces the result in the same slot. %
While this approach is straightforward and correct, it is unsuitable for an optimizing compiler, as it does not set the program up for further simplification.
For example, in the case of a loop \texttt{for i in 0..10: z[i]=x[i]+y[i]}), each iteration would result in a unique operation with distinct operands, each rotated by different amounts.

	Instead, \oursystem introduces the notion of a \emph{target slot}, determined by the further uses of the result.
	For example, if the result of a computation is assigned to \texttt{z[k]}, we select $k$ as the target slot, eliminating the rotation required afterward.
	If no clear target slot can be derived from the uses of the result, we use one of the operand indices as the target slot, removing the need to rotate that operand.
	Selecting the target slot this way reduces the immediate number of rotation operations created.
	More importantly, it reliably maps operations with the same \emph{relative} index access patterns to the same set of rotations and SIMD operations, allowing them to be eliminated by  
	 the following simplification pass.
	Since this approach is based purely on index access patterns, it works equally well for complex loop nests and heavily interleaved code.

\paragraph{Operand Conversion}
	In order to convert a scalar operation to a fully-batched SIMD operation, all non-batched inputs must be converted, as described in \Cref{algo:batching_pass} ({\sc{OperandConversion}}).
	Note that, due to the linear-walk nature of the pass, all previous FHE operations have already been converted to fully-batched operations.
	As a result, any operand of type \texttt{fhe.secret} must be the result of an \texttt{fhe.extract} operation.
	This invariant allows us to `chain' batched operations together by replacing the extraction-based operand with a rotation-based operand.
	Specifically, an operand extracted from slot $i$ of vector $x$, is replaced with a rotation of $x$  by $i - ts$, where $ts$ is the target slot determined before.

\paragraph{Ensuring Consistency}
	We maintain the consistency and correctness of the program at each step of the optimization.
	Towards this, we first construct the new rotations and batched operations as additions to the program.
	We only replace occurrences of the old operation with the optimized version after we 
	replace uses of the old operation with an \texttt{extract} operation that extracts the target slot of the new batched result.
	This ensures that, even if no further batching opportunities are found, the program remains correct.
	Since batched FHE schemes do not support true scalar values,
	we simply interpret scalars as ciphertexts where only slot $0$ contains valid data.
	With this convention, any remaining extractions will eventually be converted to a rotation by $-ts$.
	However, in practice, this is rare as most of the \texttt{extract} operations we insert will in turn be converted to rotations when the next homomorphic operation is processed.
	As a result, these consistency-related \texttt{extract}  operations are frequently eliminated completely at the end of the batching pass.

\vspace{-1em}
\subsection{Rotate-and-Sum Pass}
\label{folding-opt}

After the main pass and the associated simplification pass,
we apply the \emph{rotate-and-sum} technique where applicable.
Since this optimization requires a holistic view of the operation, 
this would be significant if we did not merge sequential operations during preprocessing.
While we used a sum over all elements when explaining the technique in the previous section, the technique can be generalized to any subset with a consistent stride.
Additionally, it can also be used to compute products rather than sums.
When applied to multiplication, it additionally has the benefit of automatically reducing the multiplicative depth of the expression as a side-effect.

Note that the pre-processing combination of binary operations into larger operations \emph{must} happen before the main pass described above, as that pass would otherwise insert rotations between the different operations, making them no longer directly chained.
The actual translation to a series of rotations and native binary operations, meanwhile, has to be performed \textit{after} that pass, since it requires the operands to be entire ciphertexts, rather than scalars.
Additionally, the de-duplication simplifications that can take place after the batching transformation can widen the applicability of this transformation by reducing the number of distinct ciphertexts appearing in the program.

\vspace{-1em}
\section{Implementation}
We build \oursystem on top of the open-source MLIR framework~\cite{mlir}, which is rapidly establishing itself as the go-to tool for domain-specific compilers and opening up the possibility of exchanging ideas and optimizations even beyond the FHE community.
\oursystem consists of roughly 15k~LOC of C++, with around 2k~LOC of Python for the Python front-end.
\oursystem uses the Microsoft \acf{seal} as its FHE backend.
\ac{seal}, first released in 2015, is an open-source \ac{fhe} library implemented in {\cpp} that is thread-safe and heavily multi-threaded itself.
SEAL implements the \acs{bfv}, \acs{bgv} and \acs{ckks} schemes.

In contrast to existing monolithic compilers, \oursystem is highly modular and designed to be flexible and extensible.
We decouple optimizations from front-end logic, allowing for a wide variety of domain-specific front-ends and the ability to easily replace back-ends to target different FHE libraries or hardware accelerators as they become available. 
The toolchain can easily be adapted to different needs, with certain optimizations enabled or disabled as required.

\section{Evaluation}
\label{evaluation}

\oursystem is designed to compile high-level programs, written by non-experts in the standard imperative paradigm,
into highly efficient batched FHE implementations that achieve the same performance as hand-crafted implementations by experts.
\oursystem achieves its usability goals through a well-integrated Python front-end and by requiring developers to alter their code only minimally  (annotating variables as secret).
However, ease-of-use becomes moot when the performance of the generated code is not competitive.
Therefore, we focus our evaluation on the performance of \oursystem and the code it generates,
trying to answer whether or not automatic optimizations can bring naive code to the same performance level as expert implementations.
In this section, we first show the effect of the batching optimizations on benchmark workloads designed to demonstrate different batching patterns,
then compared against synthesized optimal batching patterns,
and finally discuss a real-world application example.

\begin{figure*}[!t]
    \captionsetup[subfigure]{aboveskip=2pt,belowskip=8pt}
    \centering
    \begin{subfigure}[t]{0.45\textwidth}
        \includegraphics[width=\textwidth, trim={0.1cm 0.2cm 0.1cm 0.2cm},clip]{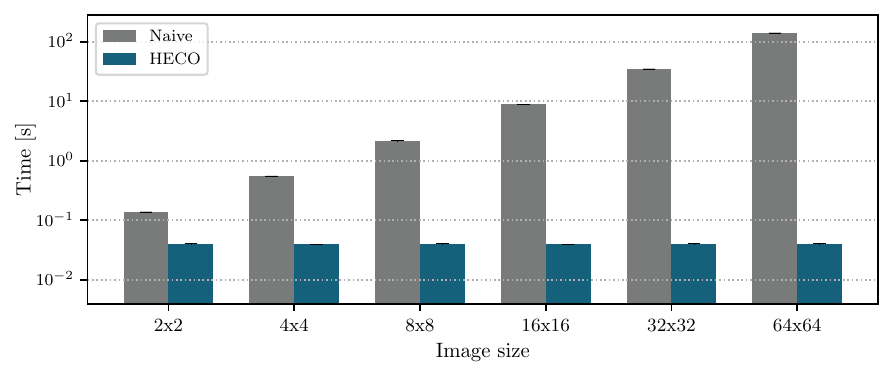}
        \subcaption{Roberts Cross benchmark runtime (in seconds).}
        \label{robertscros}
    \end{subfigure}%
    \hfill
    \begin{subfigure}[t]{0.45\textwidth}
        \includegraphics[width=\textwidth, trim={0.1cm 0.2cm 0.1cm 0.2cm},clip]{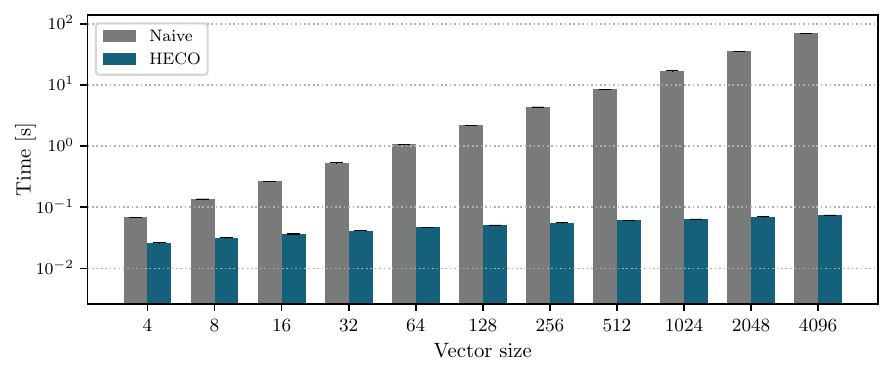}
        \subcaption{Hamming Distance benchmark runtime (in seconds).}
        \label{hammingdist_memory}
    \end{subfigure}
    \vfill
    \begin{subfigure}[t]{0.45\textwidth}
        \includegraphics[width=\textwidth, trim={0.1cm 0.2cm 0.1cm 0.2cm},clip]{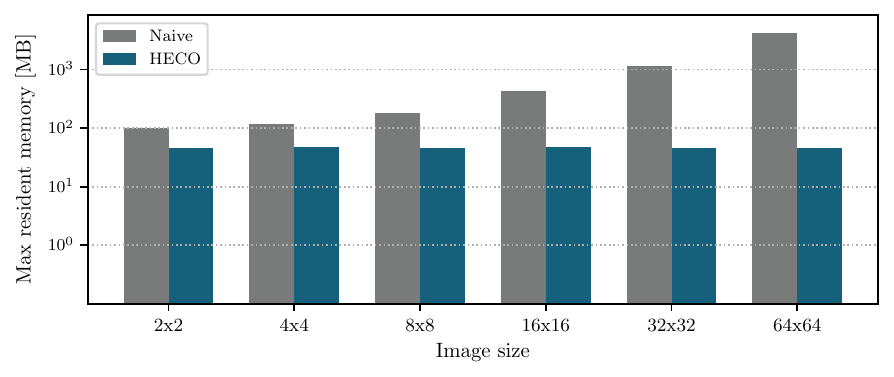}
        \subcaption{Roberts Cross benchmark memory usage (in MB).}
        \label{robertscross_memory}
    \end{subfigure}%
    \hfill
    \begin{subfigure}[t]{0.45\textwidth}
        \includegraphics[width=\textwidth, trim={0.1cm 0.2cm 0.1cm 0.2cm},clip]{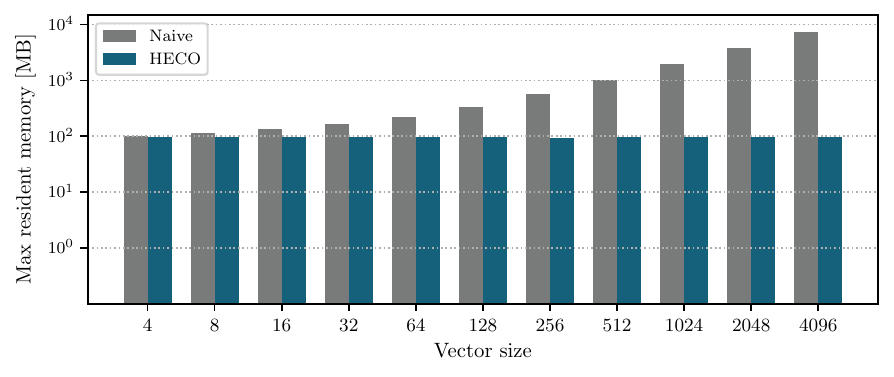}
        \subcaption{Hamming Distance benchmark memory usage (in MB).}
        \label{hammingdist}
    \end{subfigure}%
    
    \caption{Log-log plot of the runtime and memory consumption of the roberts cross and hamming distance benchmarks for different vector sizes, comparing a naive non-batched solution with the batched solution generated by our system.}
\end{figure*}

\vspace{2em}
\subsection{Benchmarks}
We evaluate \oursystem in terms of the speedup reduction in memory overhead gained over non-optimized implementations and the compile time required.

\paragraph{Applications}
We demonstrate the speedup achieved by our batching optimization on two applications that are representative of common batching opportunities.
The \emph{roberts cross operator} is an edge-detection feature used in image processing.
It approximates the gradient of an image as the square root of the sum-of-squares of two different convolutions of the image, which compute the differences between diagonally adjacent pixels.
As in all other kernel-based benchmarks,  wrap-around padding is used,  which aligns well with the cyclical rotation paradigm of FHE.
In order to enable a practical FHE evaluation, the final square root is omitted, since it would be prohibitively expensive to evaluate under encryption.
The \emph{hamming distance}, meanwhile, computes the edit distance between two vectors, i.e., the number of positions at which they disagree.
Here, we consider two binary vectors of the same length, a setting in which computing (non-)equality can be done efficiently using the arithmetic operations available in FHE.
Specifically, this makes use of the fact that $\mathrm{NEQ}(a,b) = \mathrm{XOR}(a,b) = (a-b)^2$ for $a,b \in \{0,1\}$.

\vspace{-1em}
\paragraph{Baseline}
Our baseline is a naive implementation of the application without taking advantage of batching, as one might expect FHE novices to implement.
In this setting, vectors of secrets are directly translated to vectors of ciphertexts.
While this approach introduces significant ciphertext expansion and increases the memory required, it is actually preferable in terms of run time over a solution that batches vector data into ciphertexts, but does not re-structure the program to be batching-friendly.
This is because such a solution adds the overhead of rotations, masking, etc., to the base runtime of the non-batched solution.

\vspace{-1em}
\paragraph{Environment}
All benchmarks are executed on AWS \texttt{m5n.xlarge} instances, which provide 4 cores and 16~GB of RAM.
We used Microsoft SEAL~\cite{sealcrypto} as the underlying FHE library, targeting its BFV~\cite{brakerski2012fully,fan2012somewhat} scheme implementation.
All experiments are run using the same parameters, which ensure at least 128-bit security.
We report the run time of the computation itself, omitting client-side aspects such as key generation or encryption/decryption.
All results are the average of 10 iterations, discarding top and bottom outliers.

\vspace{-1em}
\paragraph{Runtime \& Memory Overhead}
In \Cref{robertscros} we show the runtime of the Roberts Cross benchmark for varying instance sizes, comparing the non-batched baseline with the batched solution generated by \oursystem.  
While the run time of the naive version increases linearly with the image size, the batched solution maintains the same performance (until parameters must be increased to accommodate even larger images).
Instead, the run time is more closely tied to the size of the kernel than to that of the image.
This highlights the dramatic transformations achieved by \oursystem, fundamentally changing the structure of the program.
As a result of these transformations,  \oursystem achieves a speedup of 3454x over the non-batched baseline for 64x64 pixel images,  demonstrating the extraordinary impact that effective use of batching can have on FHE applications: while the non-batched solution is borderline impractical at over two minutes, the batched solution takes only a fraction of a second (0.04~s).  
The runtime of the generated code in this case does depend directly on the vector length.
However, due to the fold-style optimization (cf. \Cref{folding-opt}), this dependence is only logarithmic.

\begin{figure*}
	\def\myvspace{3pt}
	\vspace{3pt}
	\centering
	\hspace{-1em}
	\includegraphics[width=\textwidth]{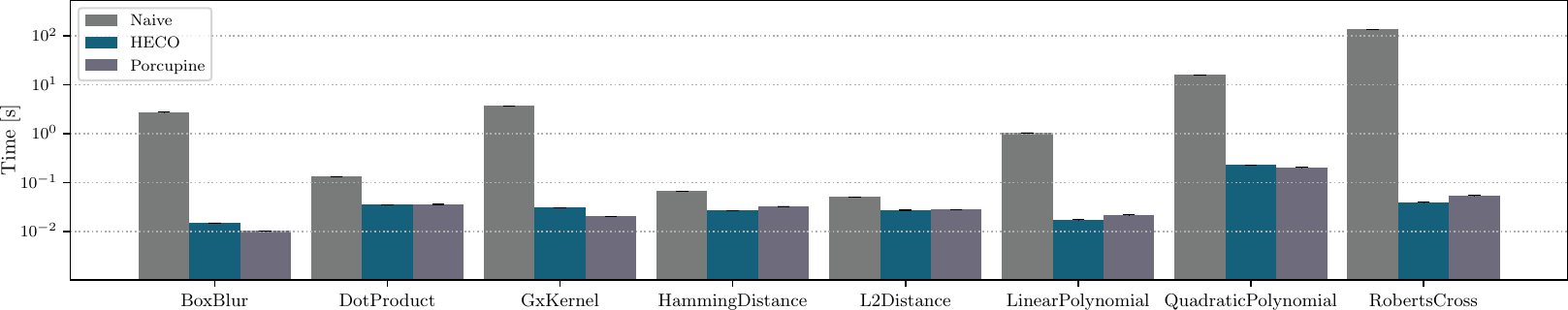}
	\caption{\scalecaption Runtime of example applications (in seconds), comparing a naive non-batched baseline, the solution generated by our system (\oursystem), and an optimally-batched solution synthesized by the Porcupine tool.}
	\label{porcupine}
\end{figure*}

In \Cref{hammingdist} we can see that, for realistic problem sizes, the performance advantage of batching becomes significant, resulting in a speedup of 934x for 4096-element vectors.
We also see that, while the runtime of the batched solution does increase with the vector length, this is nearly imperceptible when compared to the non-batched baseline.
Finally, in \Cref{robertscross_memory}/\ref{hammingdist_memory},
we show the memory overhead of the non-batched baseline and the batched solution.
While FHE introduces a non-negligible baseline overhead due to the large amount of key- and other context-data that must be maintained, the reduced number of ciphertexts in the batched solution has a clear impact on memory usage,
increasingly so as the problem sizes increase.

\paragraph{Compile Time}
\oursystem achieves these fundamental transformations efficiently, with compile times that are amenable to interactive development.
This is in contrast to synthesis-based tools, which
 require more than 10~minutes to synthesize a batched solution for the Roberts Cross benchmark  even for toy-sized instances and do not scale to the sizes we consider here at all~\cite{Cowan2021-vx}.

\subsection{Comparison with Synthesized Solutions}
We compare the baseline and \oursystem to synthesized optimal batching patterns.
Synthesis based approaches explore the space of all possible programs, constrained by a reference specification describing input-output behavior.
While this tends to be computationally expensive, it has the potential to find optimal solutions featuring highly non-intuitive optimizations.
We use a set of nine benchmark applications that represent a variety of common code patterns relevant to batching for our evaluation.
These programs range from having no inter-element dependencies (e.g., Linear Polynomial), to simple accumulator patterns (e.g., Hamming Distance), and complex dependencies across a multitude of different vector elements (e.g., Roberts Cross).
As a result, they provide a useful benchmark, especially for batching optimizations.
These benchmarks were first proposed in~\cite{Cowan2021-vx}, which introduces Porcupine~\cite{Cowan2021-vx}, a synthesis-based compiler for batched FHE. 
In addition to a reference specification, it requires a developer-provided sketch of an initial possible batched approach. 
Our `synthesis' solutions are based on pseudo-code made available in an extended version of the paper~\cite{Cowan2021-vx}.

Synthesis based tools can require the significant search time to find solutions, limiting them to toy-sized workloads. For example, Porcupine requires over 10 minutes to synthesize a program with ten instructions and will fail to synthesize a solution at all for sufficiently complex programs.
As a result, we consider the following problem sizes here:
The synthesized dot-product code targets 8-element vectors, while those for Hamming Distance and L2 Distance were provided for 4-element vectors.
For the other applications, we use vectors of length 4096, representing 64x64 pixel images

As we can see from \Cref{porcupine}, \oursystem dramatically improves performance over the non-batched baseline approach.
For example, for the Roberts Cross benchmark, our batched solution is over 3500 times faster than the non-batched solution, taking less than 0.04~seconds instead of over  2.35~minutes.
More importantly, our results are nearly equivalent to the optimally batched solutions synthesized by Porcupine, especially when considering the stark contrast between the non-batched and the two batched solutions.
In some cases (e.g., Box Blur), Porcupine has an advantage because it finds non-intuitive solutions beyond traditional batching patterns.
Interestingly, for some applications (e.g. Hamming Distance) \oursystem actually outperforms Porcupine.
This is because Porcupine provides an optimally batched solution but does not necessarily handle ciphertext management optimally, inserting unnecessary relinearization operations.

\begin{table}[t]
    \begin{tabularx}{\columnwidth}{XXXXXXX}
        \toprule
        {$n$}         & {4}  & {16} & {64} & {256} & {1024} & {4096} \\ \midrule
        \textbf{rc}   & 0.06 & 0.07 & 0.08 & 0.12  & 0.30   & 1.28   \\
        {\textbf{hd}} & 0.03 & 0.04 & 0.06 & 0.09  & 0.25   & 1.85   \\
        \bottomrule
    \end{tabularx}
    \caption{Compile time (in seconds) of the Roberts Cross (rc) and Hamming Distance (hd) benchmarks for different problem sizes ($n$).}
    \label{compile-times}
\end{table}

\vspace{-1em}
\subsection{Real-World Application}
The previous benchmarks demonstrated \oursystem's effectiveness and performance for different and common batching patterns.
We now evaluate an application that more closely resembles the complexity that real-world settings exhibit.
Specifically, we consider an application computing private statistics over two databases that might contain duplicate entries.
We use this to demonstrate that \oursystem can produce efficient FHE code for non-trivial programs, 
while also highlighting that there is further room for optimizations exploiting application semantics.

\paragraph{Application}
	Privacy regulations frequently prohibit entities from combining sensitive datesets directly.
	Instead, they could employ \emph{threshold} FHE, which extends FHE with multi-party key generation, to securely compute on the (encrypted) joint dataset.
	In this setting, neither party has sole access to the secret key, and they must collaborate to decrypt the results of approved queries.
	However, in practice their datasets might \emph{overlap} (e.g., agencies at different levels of government collecting similar data), introducing duplicate items into the joint dataset.
	Due to the duplicates, analytics (e.g., counting queries) will return incorrect results.
	Therefore, we must first de-duplicate the encrypted databases before executing the analytics.
	Since threshold FHE does not affect the server-side execution of the computation, \oursystem can be used directly to develop such an application.
	This first computes the \emph{Private Set Union (PSU)}
	of two databases A,B indexed by unique IDs consistent across both (e.g., a national identifier like the SSN).
	For simplicity, we consider databases with one data column and a simple \texttt{SUM} aggregation.
	However, the presented approach trivially extends to larger databases and more complex statistics.
	\Cref{listing:psu} shows the application expressed using the \oursystem Python frontend, for a database size of $128$ elements and $8$-bit identifiers. 
	We split the identifiers into individual bits, allowing us to compute the equality function even while working with arithmetic circuits. %
	The program begins by aggregating A's data and then proceeds to check each element of B's database for potential duplicates in A.
	In order to compute the equality function, we compute $\bigwedge_k  a_k \oplus b_k$,
	using the fact that, for inputs $a,b \in \{0,1\}$, \texttt{xor} can be computed as $(a-b)^2$, \texttt{and} directly via multiplication, and \texttt{not} as $1-a$.
	If a duplicate is found, the element of B is multiplied with \texttt{unique == 0}, i.e., it does not contribute to the overall statistics.

\begin{lstfloat}[tbp]
	\centering
\begin{lstlisting}
def encryptedPSU(a_id: Tensor[128,8,Secret[int]],
                 a_data: Tensor[128,Secret[int]],
                 b_id: Tensor[128,8,Secret[int]],
                 b_data: Tensor[128,Secret[int]])
                 -> Secret[int]:
  sum: Secret[int] = 0
  for i in range(0, 128):
    sum = sum + a_data[i]
      for i in range(0, 128):
        unique: Secret[int] = 1
          for j in range(0, 128):
	    # compute a_id[i] /= b_id[j]
		eq: sf64 = 1
		for k in range(0, 8):
		   # a xor b == (a-b)^2
		   x = (a_id[i][k] - b_id[j][k])**2
		   nx = 1 - x  # not x
		   eq = eq * nx # eq and nx
		   neq = 1 - eq # not eq
		   unique = unique * nequal
		
  sum = sum + unique * a_data[i]
  return sum\end{lstlisting}
	\vspace{-3em}
	\caption{Computing statistics over duplicated data.}
	\label{listing:psu}
\end{lstfloat}
 
 \paragraph{Performance \& Discussion}
 We evaluated both a naive baseline and the \oursystem-optimized batched implementation using the same setup described earlier in this section.
 The naive approach has a run time of several minutes ($11.3~min$) and requires the sending of over 2000 ciphertexts between the server and the clients.
 The batched solution produced by our system requires not only significantly less data to be transmitted (only 4 ciphertexts), but also runs an order of magnitude faster ($57.6~s$), 
 confirming the trend we observed when evaluating on smaller benchmarks.
 While the results achieved by \oursystem are more than practical already,
 a state-of-the art hand-written implementation designed for this task can improve this even further, requiring only $1.4~s$.
 However, arriving at this solution requires a significant rethinking of the program and an application-specific batching pattern, which \oursystem intentionally does not consider to avoid a search space explosion.
 Note that synthesis based tools such as Porcupine also cannot capture these kinds of transformations.
 
 Specifically, instead of batching the identifiers for each database into a single ciphertext,
 the expert solution instead creates one ciphertext per bit, using significantly oversized ciphertexts with $128^2 = 2^{14}$ slots.
 This enables the expert solution to batch every possible permutation of the identifier set into one ciphertext. 
 By applying this to the encryption of set B, while simply encrypting $128$ non-permuted repetitions of set A, the expensive $\O(n^2)$ duplication check can be performed in parallel on all elements at the same time. 
 Computing the \texttt{unique} flag then uses a rare application of the rotate-and-\emph{multiply} pattern.
 As a trade-off, the equality computation is no longer batched, but since the number of bits in the identifiers is, by necessity, at most logarithmic in the number of database elements, this is a profitable trade-off.

 In general, exploiting application-specific packing patterns can unlock additional performance gains and is a frequently combined with client-side processing in expert-designed FHE systems.
 However, the decision on which client-side processing (e.g., removing outliers, computing permutations, etc)  is sensible is not a well-defined problem that automated solutions can tackle. 
 At the same time, the performance improvements demonstrated by \oursystem provide a first jump from the regime of prohibitive overheads to one of practical solutions.
 While further optimizations are likely frequently possible, they offer quickly diminishing returns.
 For example, scaling this application to real-world sizes (which, incidentally, is more complex with the expert approach) means that a naive solution might take days to compute while \oursystem's solution would complete in a few hours, which is a reasonable runtime for these kinds of secure statistics.

\section{Related Work} \label{relwork}

In this section, we briefly discuss related work in the domain of \ac{fhe} compilation (\Cref{relwork:fhe_compiler}) followed by a discussion of differences to existing \ac{mpc} and \ac{zkp} compilers (\Cref{relwork:other_compilers}).

\subsection{\ac{fhe} Compilers} \label{relwork:fhe_compiler}
The complexity of implementing FHE operations efficiently led to the development of dedicated libraries~\cite{Halevi2013,sealcrypto} early on.
Today, a large number of libraries provide efficient implementations of state-of-the-art schemes.
These libraries mostly provide comparatively low-level APIs that allow developers to extract the best possible performance but require significant expertise to utilize effectively.
As a result, a first wave of FHE tools and compilers emerged that tried to improve the usability of FHE~\cite{viand2021sok,Viand2018-cs,Carpov2015-pi,Chielle2020-qh,Dathathri2019-vu}.
These mostly target a circuit-level abstraction and are focused on circuit optimizations~\cite{viand2021sok}.

For example, Microsoft's EVA~\cite{Dathathri2019-vu} offers a user-friendly high-level interface and automatically inserts ciphertext maintenance operations into the circuit.
EVA uses a custom circuit-based IR and requires developers to manually map their program to the FHE programming paradigm.
In order to ease this process, recent versions~\cite{Chowdhary2021-go} include a library of expert-implemented batched kernels for frequently used patterns, e.g., summing all elements in a vector.
However, this still requires developers to manually transform an application to the batched paradigm.

A series of tools including Cingulata~\cite{Carpov2015-ok}, E3~\cite{Chielle2020-qh}, SyFER-MLIR~\cite{Govindarajan-sm}, and Google's Transpiler~\cite{google-transpiler} attempt to  translate arbitrary programs without the usual restrictions of FHE.
They achieve this by translating their input programs into binary circuits, encrypting each input bit individually.
However, programs translated in this way are virtually always too inefficient to be of practical use because they do not support the type of  high-level transformations that \oursystem employs to achieve practical efficiency.

Domain-specific compilers~\cite{Dathathri2019-ha, Boemer2019-mt, boemer2019}, e.g., targeting encrypted Machine Learning applications, rely on a large set of hand-written expert-optimized kernels for common functionality (mostly linear algebra operations).
Since these tools rely on pre-determined mappings rather than automatically identifying optimization opportunities, they do not transfer to other domains, such as the general-purpose setting \oursystem targets.
Besides that, their lack of flexibility prevents their use when developers' needs are even slightly misaligned.

The Porcupine compiler~\cite{Cowan2021-vx} is closest to our work in that it also considers translating imperative programs to FHE's batching paradigm.
However, their tool has a significantly different focus, using a heavy-weight synthesis approach that tries to identify optimal solutions that can outperform even state-of-the-art approaches used by experts. 
Since it explores a large state space in the search for an optimal solution, compile times tend to be long (up to many minutes) and programs can contain at most a handful of statements before the approach becomes infeasible.
Additionally, Porcupine requires that developers provide a sketch of the structure of the batched program, making it less suitable for non-expert users.

Finally, we want to highlight that the MLIR framework is rapidly establishing itself as the gold standard for FHE tooling.
Early attempts such as SyFER-MLIR~\cite{Govindarajan-sm} relied primarily on built-in optimizations, adding only a few binary-circuit-based FHE-specific rewrite rules.
No evaluation is provided for SyFER-MLIR, but prior work studying similar simple rewrite rule-based tools~\cite{viand2021sok} leads us to predict that it would produce only relatively minor speedups.
More recently, however, a variety of concurrent work has successfully realized different aspects of the FHE ecosystem using MLIR.
For example, Zama's Concrete-ML ~\cite{Zama_undated-ie} internally uses an MLIR-based compiler to translate Machine Learning tasks expressed as Numpy programs to the TFHE scheme. 
HECATE~\cite{Lee2022-nh}, meanwhile, improves upon the rescale-allocation optimizations presented in the EVA compiler~\cite{Dathathri2019-vu}.
However, where the latter uses a custom Python implementation that does not provide for interoperability with other tools, HECATE builds upon common MLIR abstractions.
Thanks to the modular nature of MLIR, these tools could easily be integrated into HECO's end-to-end toolchain.

\subsection{\ac{mpc} \& \ac{zkp} Compilers} \label{relwork:other_compilers}
Compilers for both Multi-Party Computation (MPC) and Zero-Knowledge Proof (ZKP) systems 
face similar challenges to FHE compilation, such as the need for data-independent computations and a general tendency towards trying to achieve small and low-depth circuits.
However, in practice we find these similarities are too superficial to allow techniques from one domain to be lifted to another.
Due to the heavy reliance on selectively revealing (potentially blinded) data during an MPC computation -- a feature that has no direct correspondence in FHE -- many of the optimization approaches are unlikely to transfer.
State-of-the art MPC compilers~\cite{Buscher2017-zh, Hastings2019-qw} also frequently make heavy use of hybrid approaches, i.e., switching between different MPC settings. 
While there has been significant work on scheme switching for FHE~\cite{pegasus,Boura2018-wj}, practical applications remain rare and few libraries currently support these techniques. 
As these approaches start to mature, investigating to what extent scheme-switching optimizations from the domain of MPC can transfer to FHE will present an interesting avenue for future work.   

Zero-Knowledge Proof Compilers also face the challenge of mapping complex operations to arithmetic circuits with limited expressiveness.
However, their setting fundamentally differs from that of FHE, as the prover generally has access to all data in the computation in the clear.
This allows ZKP computations to heavily rely on witness-based computation, which allows compilers to shift virtually all non-arithmetic operations outside the core ZKP computation.
Additionally, ZKP compilers mostly use intermediate representations based on Rank-1 Constraint Systems (R1CS) or other constraint specification systems that are not suitable for expressing FHE computations.
Finally, we note that compiler frameworks trying to accommodate MPC, ZKP and potentially also FHE are emerging~\cite{Ozdemir2022-fx}.
However, their reliance on a circuit-like IR makes them unsuitable for the high-level transformations we use in our work.

\section{Discussion}
	As FHE has emerged into practicality, it has drawn the interest of a significantly wider audience bringing new perspectives, requirements and backgrounds to the area.
	While traditionally, FHE applications were mostly developed by the same experts that designed, optimized and implemented the underlying cryptographic schemes, 
	this will soon no longer be true beyond the world of cutting-edge academic research.
	In recent years, a variety of tools has emerged in an attempt to address the needs of future non-expert developers.
	While some domain specific tools have proven to be very effective~\cite{viand2021sok, boemer2019, Chillotti2021-qo},
	most general purpose tools have fallen short of delivering on the promise of usable FHE.
	While they simplify the development process, they generally produce naive implementations which provide little real-world benefit due to their significant overhead compared to more optimal implementations.
	However, as performance is key for practical deployment, we believe that usability without sufficient performance is mostly meaningless.

	\oursystem aims to bridge the gap between usability and performance for general-purpose workloads.
	It offers non-expert developers the ability to express applications in a familiar high-level paradigm \emph{without} paying the extreme performance penalty this would usually incur.
	Beyond optimizing this high-level transformation, \oursystem proposes a new end-to-end architecture for FHE compilers based on the distinct stages of FHE optimization we identify.
	\oursystem's modular architecture is designed to allow it to interoperate with other toolchains and easily integrate future optimization techniques.
	While \oursystem represents an important step in FHE usability, many challenges remain to be addressed.
	For example, \oursystem considers RLWE-based schemes offering SIMD operations,
	but recent developments in LWE-based fast-bootstrapping schemes makes them an attractive alternative. 
	Today, these worlds remain mostly separate and use significantly different paradigms.
	Future work needs to consider how to unify these, 
	especially	in the context of scheme-switching, i.e., the ability to move between schemes inside a single application.
	Finally, upcoming dedicated FHE hardware accelerators promise to deliver significant performance improvements but require sophisticated scheduling to unlock their potential.
	While existing work on accelerators already incorporates automated scheduling, there are likely significant further optimization opportunities in considering compilation for these systems from an end-to-end perspective.
	More generally, we believe that there is significant potential for interdisciplinary research that combines techniques from compiler and programming language research with insights from cryptography.

 \section*{Acknowledgments}
We thank Gyorgy Rethy and Nicolas Küchler for their invaluable help in evaluating HECO. 
We also thank our anonymous reviewers, Tobias Gross, Michael Steiner, and the PPS Lab team for their insightful input and feedback.
We would also like to acknowledge our sponsors for their generous support, including Meta, Google, SNSF through an Ambizione Grant No. 186050, and the Semiconductor Research Corporation.

\makeatletter 
\bibliographystyle{ieeetran}
\interlinepenalty=10000
\bibliography{references}

\begin{thebibliography}{10}
\bibitem{Zama_undated-ie}
{Zama}, ``concrete-numpy: Concrete numpy is an open-source library which
  simplifies the use of fully homomorphic encryption.''
\end{thebibliography}
\makeatother

\end{document}